\documentclass[twocolumn,english,preprintnumbers,amsmath,amssymb,prb]{revtex4}
                                                                                   
\usepackage{graphicx}
                                                                                   
\begin{document}

\title{Hydrogen dynamics on defective monolayer graphene}
\author{Carlos P. Herrero}
\author{Jos\'e A. Verg\'es}
\author{Rafael Ram\'irez}
\affiliation{Instituto de Ciencia de Materiales de Madrid (ICMM),
         Consejo Superior de Investigaciones Cient\'ificas (CSIC),
         Campus de Cantoblanco, 28049 Madrid, Spain }
\date{\today}

\begin{abstract}
The hydrogen dynamics on a graphene sheet is studied
in the presence of carbon vacancies.
We analyze the motion of atomic H by means of molecular
dynamics (MD) simulations, using a tight-binding Hamiltonian
fitted to density-functional calculations.
Hydrogen passivates the dangling bonds of C atoms close to
a vacancy, forming C--H bonds with H located
at one or the other side of the layer plane.
The hydrogen dynamics has been studied from statistical analysis
of MD trajectories, along with the autocorrelation function
of the atomic coordinates.
For a single H atom, we find an effective barrier of 0.40~eV
for crossing the graphene layer, with a jump rate
$\nu = 2 \times 10^6$~s$^{-1}$ at 300~K. The atomic jumps behave
as stochastic events, and their number for a given temperature
and time interval follows a Poisson probability distribution.
For two H atoms close to a vacancy, strong correlations in
the atomic dynamics are found, with a lower jump frequency
$\nu = 7 \times 10^2$~s$^{-1}$ at room temperature.
These results provide insight into the diffusion mechanisms
of hydrogen on graphene, paving the way for a complete
understanding of its motion through defective crystalline
membranes.  \\

\end{abstract}

\maketitle

\section{Introduction}

The study of hydrogen as an impurity in solids and on surfaces 
dates back to many years.
Although it is one of the simplest impurities, a deep understanding
of the physical and chemical properties of H-related defects 
is not obvious because of its low mass, 
and needs the application of advanced  theoretical and 
experimental techniques.\cite{pe92,es95}
For two-dimensional (2D) systems, such as graphene, hydrogen
chemisorption can nowadays be efficiently obtained in the laboratory 
by several routes.\cite{wh18} The resulting material may be suitable 
for catalytic reactions,\cite{fa15,hu17} and turns out to be
a good candidate for hydrogen storage.\cite{di01,to13,ka21,su21}

Experimental and theoretical work indicates that a defect-free
graphene sheet is impermeable to gases, even to small atoms
such as He or H.\cite{mi13,ts14,su20b}
For atomic H, in particular, the energy barrier for crossing
a perfect graphene layer is about 4~eV,\cite{ts14,gu18b} which
makes permeation highly unlikely. Sun {\em et al.}\cite{su20b}
have recently suggested that the effective barrier for this
dynamic process could be lower, thus favoring jumping
of atomic hydrogen from one side of the sheet to the other.
Proton permeation across graphene and related materials 
under various conditions turns out to be more efficient 
than in the case of atomic hydrogen.
This has been demonstrated in recent years by several 
research groups.\cite{kr17,po18b,ma18b,ba19,xu19,gr20}

Permeation of atoms and molecules through crystalline
membranes such as graphene is enhanced by the presence
of nanomeshes, nanopores, and atomic-scale
defects.\cite{el18,sc20,li21}
This has attracted interest for its applications
in gas separation\cite{li15,su19c,ti21} and sieving
of H isotopes.\cite{lo16b,lo17c}
In connection with this, hydrogen is known to passivate
vacancy defects and edges in
graphene nanoribbons.\cite{lu09b,is14,pa14b}

The properties of atomic and molecular hydrogen on graphene have 
been investigated by various research groups from a theoretical 
point of view, using {\em ab-initio} electronic structure 
methods,\cite{sl03,ho06b,ca09,bo08,du04,an08,de13,ts14,bo18b}
as well as semi-empirical potentials.\cite{pe13b,pe18b}
In particular, introducing hydrogen molecules on the graphene 
surface, the permeation process is thought to consist of several 
steps, namely molecular dissociation, atomic diffusion on the
surface, crossing of the layer, recombination, and
desorption.\cite{de13,ts14}
The actual disposition of hydrogen atoms on graphene has been
also studied by employing experimental techniques, such as scanning 
tunneling microscopy, electron diffraction, and photoemission 
spectroscopy.\cite{ba09b,ba10,ul13}

In this paper, we study the motion of atomic hydrogen in defective
graphene. In particular, we consider one and two H atoms close to a
carbon vacancy and study their dynamics, with especial emphasis
on the jump rate of hydrogen for crossing the graphene layer
from one side to the other.
We present results of tight-binding (TB) molecular dynamics (MD)
simulations in a temperature range from 300 to 1200~K, 
obtained from a TB Hamiltonian
employed earlier to study H diffusion in graphite,\cite{he10b} 
graphene,\cite{he09a} and diamond.\cite{he07}
In the case of two H atoms,
we find a strong correlation in the jumps for both of them.
As a check of the reliability of the TB Hamiltonian to describe
C--C and C--H interactions in hydrogenated graphene, we have also
carried out {\em ab-initio} density-functional theory (DFT)
calculations to determine some relevant
features of the energy surface, such as energy barriers.
TB MD simulations similar to those presented here have been
carried out earlier to study several finite-temperature properties
of hydrogen-related defects
in various types of materials,\cite{pa94,be00,bo94,sh10,uk10}
as well as on surfaces.\cite{ha11,do18,ma18c,do19}

The paper is organized as follows. In Sec~II we present the computational
methods employed here, i.e., tight-binding molecular dynamics,
DFT method, and harmonic linear-response procedure. 
In Secs~III and IV, we give results for H impurities at $T = 0$ 
and finite temperatures, respectively.
The dynamics of the hydrogenic complexes with one and two H atoms
is presented in Secs~V and VI.
The paper closes with a summary in Sec.~VII.

\section{Computational methods}

In this Section we present the methods employed in our calculations.
In Sec.~II.A we focus on molecular dynamics simulations and the
tight-binding method used along the paper to describe the interatomic
interactions.
In Sec.~II.B we briefly outline some aspects of the DFT calculations
which we carried out to compare with the TB results at $T = 0$.
In Sec.~II.C we give a short account of the harmonic linear-response
method used to obtain vibrational frequencies at finite temperatures.

\subsection{Tight-binding molecular dynamics}

We study equilibrium and dynamical properties of hydrogen in defective
graphene by means of MD simulations.
An important point in the MD procedure is a precise definition of 
interatomic interactions, which should be taken as realistic as possible.
Using {\em ab-initio} density functional or Hartree-Fock based
self-consistent potentials would largely reduce the length of
the simulation trajectories required for a reasonable statistics of
the considered variables.
Thus, we derive the interatomic forces from an efficient tight-binding 
Hamiltonian, developed on the basis of
density functional calculations.\cite{po95}

The capacity of this kind of TB methods to describe various properties of
molecular systems and condensed matter has been discussed by 
Goringe {\em et al.}\cite{go97} and Colombo.\cite{co05}
The TB Hamiltonian employed here\cite{po95}
has been found earlier to reliably describe C--H interactions in 
carbon-based materials.\cite{he06,he07} 
As an example, for a methane molecule, it yields for $A_1$ and 
$T_2$ vibrational modes in a harmonic approximation ($HA$)
frequencies of 3100 and 3242 cm$^{-1}$, respectively.\cite{he06}
The agreement with experimental frequencies\cite{jo93} of 2917 and 
3019 cm$^{-1}$  can be considered to be good, taking into 
account the usual redshift of these modes due to anharmonicity.
This TB Hamiltonian has been used in earlier work to describe
C--H interactions in diamond,\cite{he06,he07} 
graphite,\cite{he09a,he10,he10b} and graphene.\cite{he09a}

In this paper, TB MD simulations have been performed in the 
isothermal-isobaric ($NPT$) ensemble for a graphene rectangular 
supercell, containing 95 C and $n$ H atoms, with $n$ = 1 or 2.
Periodic boundary conditions were assumed in the layer plane
($x$, $y$ coordinates) and free boundary conditions have been considered
in the perpendicular $z$ direction. 
Chains of four Nos\'e-Hoover thermostats were coupled to each atomic 
degree of freedom to give the required temperature.\cite{tu98}
Another chain of four thermostats was coupled to the barostat that
controls the in-plane area of the simulation cell ($xy$ plane),
yielding a constant pressure $P = 0$.\cite{tu98,al87}
The equations of motion have been integrated by using the reversible 
reference system propagator algorithm (RESPA), which allows
us to define different time steps for the integration of the fast 
and slow degrees of freedom.\cite{ma96}
The time step $\Delta t$ corresponding to TB derived forces was 
taken as 0.5 fs, which yielded a good precision for the variables 
and temperatures considered here.
For fast dynamical variables as the thermostats, we employed a
time step $\delta t = \Delta t/4$.

\begin{figure}
\vspace{-0mm}
\includegraphics[width=7cm]{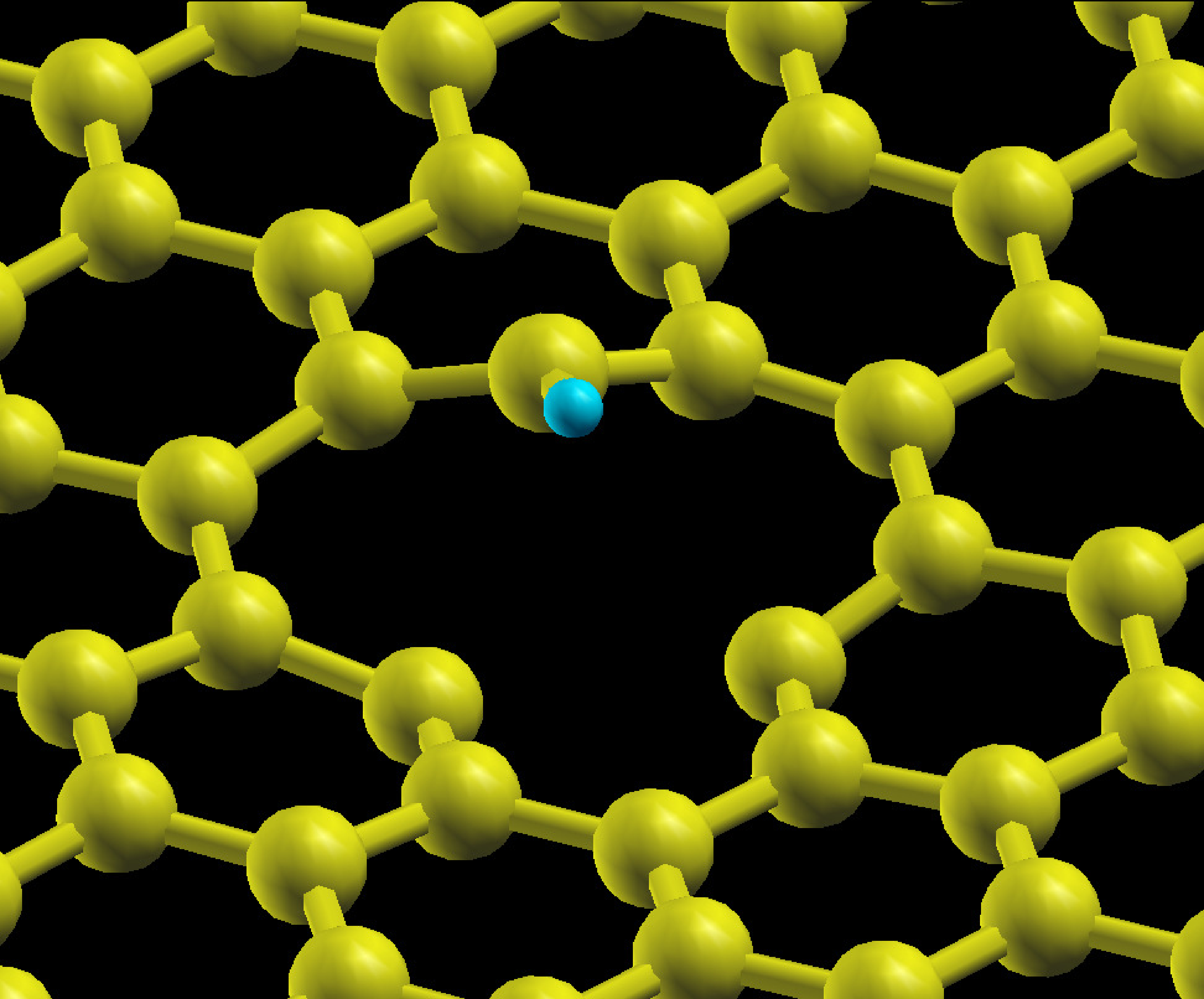}
\vspace{-0mm}
\caption{
Ball-and-stick image of an atomic configuration taken from
a MD simulation of H close to a carbon vacancy in graphene.
Yellow balls represent C atoms, whereas the small blue sphere
represents the hydrogen atom.
The simulation was carried out at $T$ = 300 K.
}
\label{f1}
\end{figure}

The configuration space has been sampled at temperatures in the range 
from 300 to 1200 K.
 For a given temperature, a typical run consisted of $2 \times 10^5$ MD
steps for system equilibration, followed by $2 \times 10^7$ steps (10 ns) 
for calculating average variables and analyzing the atom dynamics.
In Fig.~1 we present a snapshot of an atomic configuration taken
from a MD simulation at $T$ = 300 K, showing a hydrogen atom 
(small blue sphere) bound to a carbon close to a vacancy in graphene.

\subsection{DFT calculations}

To check the precision of our TB procedure to describe hydrogenated
defective graphene, we have carried out some state-of-the-art
DFT calculations.
We have calculated the adsorption energy and barriers for atomic
hydrogen on a cluster of size similar to the supercell employed
for the TB MD simulations.
Dangling bonds of carbon atoms at the cluster borders were
saturated by hydrogen atoms, whose $z$ coordinate was kept fixed
on the graphene layer plane ($z$ = 0)
to have a reference for the atomic positions at the hydrogenic defect.

DFT calculations have been done with the Gaussian16 (Revision A.03)
package\cite{ga16} using the B3LYP hybrid functional\cite{be93}
and Def2SV basis set.\cite{we05} The Berny algorithm was employed
for both minimizations (optimizations to a local minimum) and
optimizations to transition states.\cite{li06}

DFT-based {\em ab-initio} calculations have been carried out earlier
to study various characteristics of atomic hydrogen on graphene, such
as trapping of electrons,\cite{ve10} defect-induced magnetism,\cite{ya07,bo08}
and opening of a gap in the electronic density of states.\cite{du04}
This computational procedure has been also applied to study the 
chemical activity of H on graphene,\cite{an08} as well as
the effect of an applied stress on hydrogen chemisorption.\cite{mc10}
In our present context, DFT calculations have been applied to 
analyze the permeability of graphene for atomic species,
with especial emphasis on hydrogen.\cite{ts14}

\subsection{Vibrational frequencies}

The harmonic vibrational frequencies for defective graphene with 
a carbon vacancy and a hydrogen atom were obtained by diagonalization 
of the dynamical matrix for the same simulation cell 
employed for the MD simulations.
Interatomic force constants for the TB potential were calculated 
by numerical differentiation of the forces, using atom displacements 
of $10^{-4}$~\AA\ from the equilibrium positions.
This gives us a $HA$ for the vibrational modes.

To estimate anharmonic shifts of vibrational modes at finite
temperatures we have used the so-called harmonic linear response (HLR)
method, which was derived by considering the statistical mechanics of
linear response within the framework of equilibrium quantum path-integral
simulations.\citep{ra01,ra19}
In the classical limit, the HLR approach is
closely related to methods that analyze vibrational modes by studying
spatial correlations of nuclear coordinates.\cite{wh03,sc04}
In our procedure, vibrational frequencies are derived from the eigenvalues 
obtained by diagonalization of the covariance matrix of atomic 
displacements in thermal equilibrium.
This matrix is readily calculated from spatial trajectories in the
configuration space sampled by either MD or MC simulations.\cite{ra20}
Quantum vibrational energies of molecules containing C-H bonds
and several crystals have been investigated so far 
by the HLR method.\cite{lo03,he06,he10}

\section{Zero-temperature results}

In this paper we focus on hydrogen in defective graphene. For 
comparison, we give some results obtained earlier for H on
pristine graphene, in the absence of defects.
In this case, the lowest-energy configuration corresponds to H 
bound to a C atom.  This carbon 
atom relaxes out of the layer plane by 0.46~\AA, and 
the C--H bond direction is perpendicular to the layer,
with an interatomic length $d_{\rm C-H}$ = 1.17 \AA.\cite{he09a}
This coordination is understood as the breaking of a $\pi$ bond 
and the creation of a $\sigma$ bond, thus modifying
the hybridization of the C atom from $s p^2$ to 
$s p^3$.\cite{sl03,bo08,ca09}
This chemisorption of hydrogen on pristine graphene gives
rise to a defect-induced magnetic moment.\cite{ya07,ca09,bo08}

For hydrogen close to a carbon vacancy on a graphene sheet,
we have calculated the minimum-energy configuration ($T$ = 0)
by relaxing the position of H and C atoms, with the exception
of the carbon atoms at the boundary of the simulation cell, whose
motion was restricted to $z = 0$. This gives a reference for
the motion of hydrogen in the $z$ direction, perpendicular to
the layer plane.
We found a lowest-energy arrangement with H bound to
one of the three carbon atoms around the vacancy, and
located off-plane with $z_{\rm H}$ = 0.77 \AA\ 
(we take $z = 0$ for the graphene sheet).  The C atom bound 
to H also moves off-plane to $z_{\rm C}$ = 0.40 \AA, 
and the C--H bond distance results to be 1.11 \AA.
Both C atoms near the vacancy and not connected to H,
slightly relax off-plane in the opposite direction to hydrogen
to $z_{\rm C} = -0.13$~\AA.
Our DFT calculations yield a desorption energy for hydrogen of
4.4 eV, which corresponds to the binding energy for a C--H bond
appearing after passivation of one dangling bond close to
the carbon vacancy. For this atom configuration in graphene, 
Casartelli {\em et al.}\cite{ca14} found from their DFT 
calculations a binding energy of 4.24~eV. 
This energy value contrasts with the much lower binding energy 
obtained for C--H bonds in defect-free graphene 
(about 0.8 eV)\cite{ho06b,ca09,bo18b}. In this case, DFT
calculations yield a small barrier of about 0.2 eV for hydrogen
adsorption, as explained in the review by 
Bonfanti {\em et al.}\cite{bo18b}. 
For a single H atom at the carbon vacancy, we find that the 
adsorption is barrierless, in agreement with Ref.~\onlinecite{ca14}.

\begin{figure}
\vspace{-3.5cm}
\includegraphics[width=9cm]{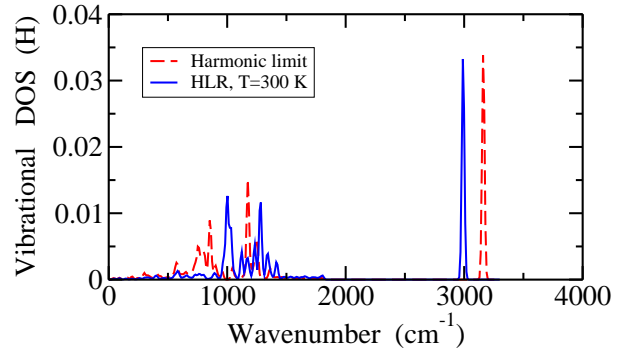}
\vspace{-0.8cm}
\caption{
Vibrational density-of-states for hydrogen in defective graphene.
The dashed line represents the DOS obtained from diagonalization of
the dynamical matrix (harmonic approximation), whereas the solid line
represents the DOS derived from HLR calculations at 300 K.
}
\label{f2}
\end{figure}

An additional characterization of this point defect in graphene 
can be obtained from the vibrational frequencies of hydrogen.
In Fig.~2 we display the vibrational density-of-states (DOS) 
for hydrogen, as derived from diagonalization of the dynamical matrix
in a $HA$ (dashed line).
The peak at 3162 cm$^{-1}$ corresponds to stretching of the C--H bond, 
which contrasts with a frequency of 2555 cm$^{-1}$ for
the C--H stretching vibration in nondefective graphene.\cite{he09a}
This is in line with a stronger C--H bond for hydrogen close to a carbon 
vacancy, as it passivates a ``dangling bond'' in this point defect, and
is also confirmed by a shorter C--H bond length in defective graphene
($d_{\rm C-H}$ = 1.11 \AA), as compared with pure graphene 
($d_{\rm C-H}$ = 1.17 \AA).\cite{he09a}
The stretching frequency obtained here for the C--H bond in defective
graphene lies between those found for $A_1$ and $T_2$ modes of
methane in a $HA$:
3100 and 3242 cm$^{-1}$, respectively.\cite{he06}

Fig.~2 reveals a broad region between 400 and 1500~cm$^{-1}$, where
H vibrations lie in the frequency range of modes associated to 
the graphene lattice.\cite{wi04,ra19}
We observe two main peaks at 854 and 1175~cm$^{-1}$, which 
correspond to modes with H displacement perpendicular 
to the C--H bond direction.

To study the hydrogen dynamics around a carbon vacancy in graphene,
an important question is the characterization of the saddle point for
hydrogen jumps from one side of the graphene sheet 
(say $z_{\rm H} > 0$) to the opposite side ($z_{\rm H} < 0$).
For the saddle point at $z_{\rm H} = 0$, our TB calculations give 
an atomic arrangement with a C--H bond distance of 1.07~\AA,
close to that of the minimum-energy configuration.
The energy barrier for H crossing from the upper
to the lower minimum is 0.37~eV, which turns out
to be a relatively low barrier for hydrogen motion.
For comparison, DFT calculations using the same graphene supercell
yield for this process a barrier of 0.42~eV.

One can also think of H jumps from one C to a neighboring C atom 
close to the vacancy, breaking a C--H bond and building a new one.
For this process we find from our TB calculations a barrier of 
1.54~eV, clearly larger than that corresponding to up-down jumps 
for H connected to a single C atom.
DFT calculations yield for this energy barrier a value of
1.47~eV, close to the TB result.
From this relatively high barrier, we expect that this kind 
of H jumps will be rare events in our finite-temperature
MD simulations, and their occurrence is expected to be much
less likely than up-down H jumps preserving the bond of hydrogen 
with the same C atom.

\begin{figure}
\vspace{-0.1cm}
\includegraphics[width=7cm]{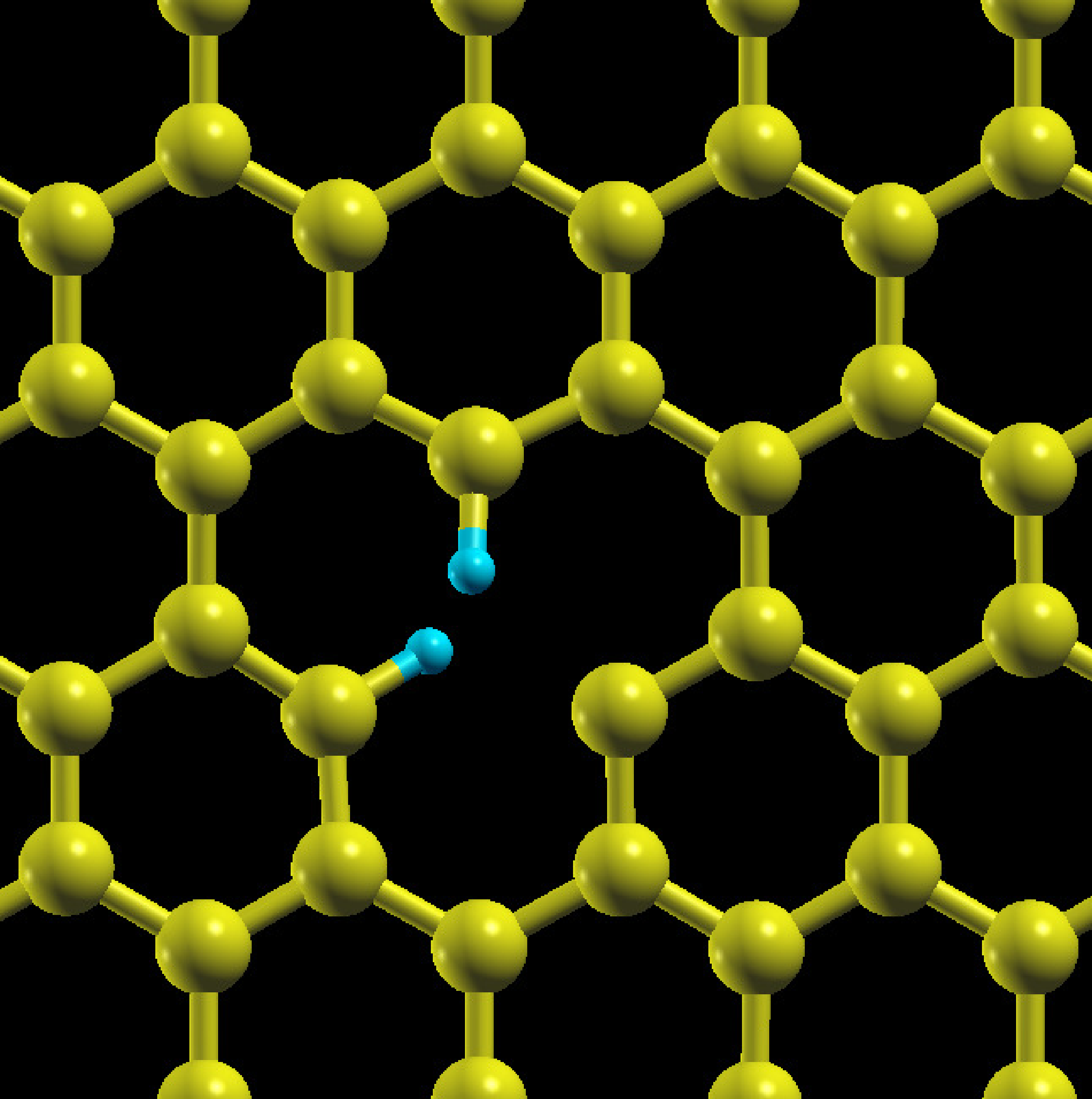}
\includegraphics[width=7cm]{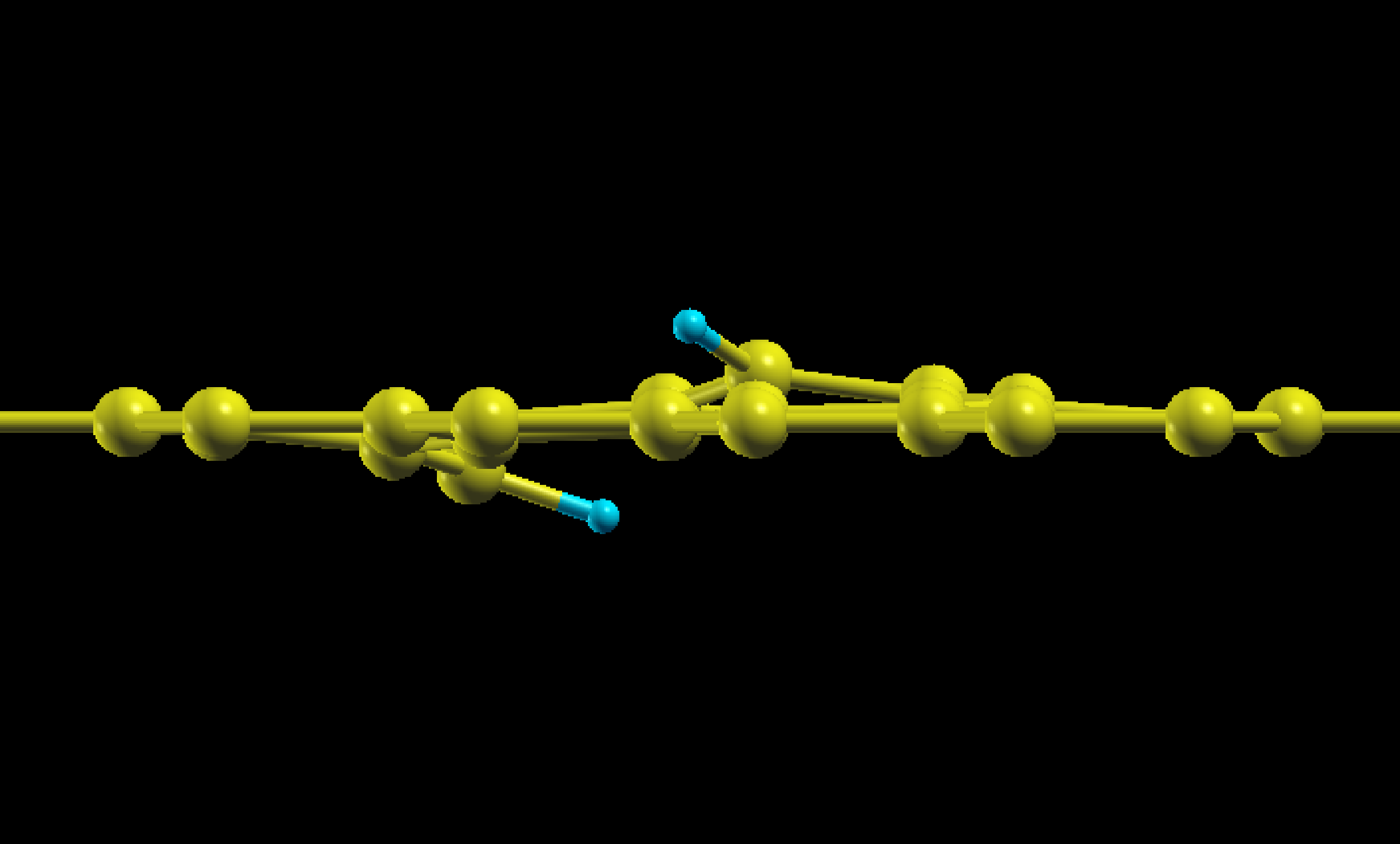}
\vspace{0cm}
\caption{
Top (upper) and side (lower) views of the minimum-energy configuration
for two H atoms at a carbon vacancy in graphene.
Yellow large balls and small blue spheres represent C and H atoms,
respectively.
}
\label{f3}
\end{figure}

We have also considered the case of two hydrogen atoms bound
to different carbon atoms near the vacancy.
In this case the minimum-energy arrangement corresponds to
one H at $z_{\rm H} = 0.75$ \AA\ and the other at 
$z_{\rm H} = -0.75$ \AA.
The C atoms bound to H are located at $z_{\rm C}$ = 0.37 and 
--0.37 \AA, respectively. 
The distance to the layer plane of both H and C 
is similar to the case of a single H atom at the vacancy, and
the C--H bond distance is 1.11~\AA.  In Fig.~3 we present 
top and side views of the minimum-energy configuration
with one H atom above the graphene layer plane and the other one
below the layer.
A configuration with both H atoms above or below the layer
plane with $|z_{\rm H}| = 1.1$~\AA\ has an energy 0.55~eV higher 
than the minimum-energy configuration. 
Such an atom disposition corresponds to a shallow local minimum 
of the energy surface, which turns out to be rather flat for
small H displacements in that region.
Thus, we have an effective repulsion between 
both H atoms, so that they prefer to stay at different sides 
of the graphene layer.

For hydrogen motion in this point defect, one can expect 
simultaneous jumps of both H atoms, interchanging its $z$ coordinate
(up-down jumps). For simultaneous crossing of the layer plane,
we obtain a relatively large energy barrier of 1.59~eV. 
This barrier is higrer than that obtained one H atom crosses
the graphene layer to a $z$ coordinate similar to the other one,
and then the second H crosses the layer plane in the opposite 
direction. This process defines the saddle-point for the cooperative
motion of both H atoms, yielding a barrier of 0.57~eV.

We have checked that the formation of molecular hydrogen
H$_2$ is unfavorable with respect to the creation of two
C--H bonds near the vacancy. In fact, for an H$_2$ molecule
plus a vacancy our TB calculations give an energy of 3.3~eV
higher than that of the most stable disposition with 
two C--H bonds (DFT yields 3.1~eV).
Another possibility is that both H atoms link to a single carbon
atom close to the vacancy, but this arrangement has an energy
1.3~eV higher than the minimum-energy configuration.

Experiemntal data of the energy barrier for atomic hydrogen 
to cross the graphene layer are scarce. This is mainly due to 
the difficulty of obtaining samples with well-characterized
defects, through which the permeation process can take place.
It has been observed that crossing a sheet of defect-free 
graphene may be enhanced by the presence of several H atoms 
clustering on the surface. Sun {\em et al.}\cite{su20b}
have reported a barrier of $1.0 \pm 0.1$~eV for atomic hydrogen,
obtained from dissociation of molecular H$_2$.
A similar barrier has been found by
Bartolomei {\em et al.}\cite{ba21} from DFT calculations.
For a specific arrangement of four neighboring H atoms in
defect-free graphene, these authors found that crossing 
the sheet by one of them is the most probable event, with 
a barrier $\Delta E$ = 0.82~eV.
These energy barriers are higher than those found here for
hydrogen flipping through the sheet in the presence of a carbon
vacancy.

\begin{figure}
\vspace{-0.5cm}
\includegraphics[width=7cm]{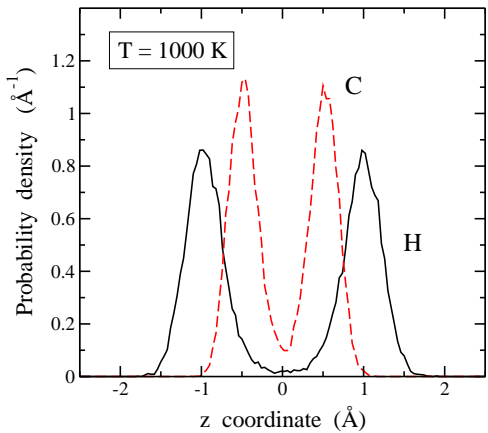}
\vspace{-0.5cm}
\caption{
Probability density for the $z$ coordinate of H (solid line)
and C bound to H (dashed line) close to a vacancy, as derived from
MD simulations at T = 1000 K.
}
\label{f4}
\end{figure}

\section{Finite-temperature configurations}

We now turn to the results of our finite-temperature simulations
for one hydrogen atom in defective graphene. 
From the calculations at $T = 0$, we have two equivalent positions 
for H bound to a C atom with $z_{\rm H}$ = 0.77 and --0.77 \AA.
In our MD simulations at finite temperatures (especially for 
$T > 400$~K), 
we observe jumps of hydrogen from one equilibrium position to
the other.  Although the highest probability density for the presence
of H occurs around these positions, we find that the mean position 
of hydrogen changes with temperature, as a consequence of thermal 
motion and anharmonicity of the C--H bond.
Thus, at $T = 500$~K we have a mean value 
$\langle |z_{\rm H}| \rangle$ = 1.01~\AA,
which slightly decreases at higher temperature to
$\langle |z_{\rm H}| \rangle$ = 0.97~\AA\ at 1200~K
(angle brackets denote an ensemble average).
Close to $|z_H|$ = 1~\AA, the potential surface for H motion is "flatter" 
for decreasing $|z_H|$, so at high temperature, 
$\langle |z_{\rm H}| \rangle$ slowly decreases.
Accordingly, the mean value $\langle |z_{\rm C}| \rangle$ 
of the C atom bound to H takes values of 0.54 and 0.50~\AA\ 
at $T$ = 500 and 1200~K, respectively.
Thus, the difference between mean $z$ coordinates of H and C atoms evolves 
from 0.37 to 0.47 \AA\ in the temperature range from $T$ = 0 to 1200~K, 
and the mean angle between the C--H bond and the layer plane increases 
from 19.5 to 25.4 degrees. 
The mean length of the C--H bond, however, changes very slowly
in this temperature range, increasing by about 0.01 \AA\ from 
the low-temperature limit up to $T$ = 1200~K.

In Fig.~4 we display the probability density of the $z$ coordinate
for H (solid line) and C (dashed line), derived from MD
simulations at $T = 1000$~K.
As expected from the results mentioned above for the mean value
$\langle |z_{\rm H}| \rangle$, the maximum probability for the 
presence of hydrogen occurs at $z_{\rm H} \approx$ 1 and -1~\AA.
For the C atom bound to H, the maxima appear at $z \approx$ 0.50
and -0.50~\AA.
Note that the density at $z = 0$ (energy barrier), even small,
is not negligible for hydrogen, $\rho_z({\rm H}) = 0.02$~\AA$^{-1}$, 
and is larger for carbon: $\rho_z({\rm C}) = 0.10$~\AA$^{-1}$.
The atomic dynamics associated to motion between both energy
minima (maxima in the probability density shown in Fig.~4)
is studied below is Sec.~V.

\begin{figure}
\vspace{-0.5cm}
\includegraphics[width=7cm]{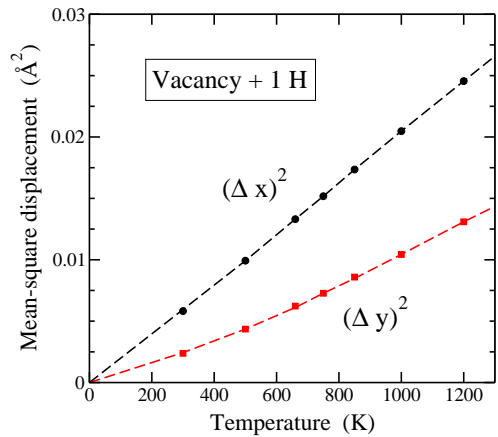}
\vspace{-0.5cm}
\caption{
Mean-square displacement of H along the $x$ and $y$ directions.
Symbols indicate results of MD simulations at several temperatures:
circles for $(\Delta x)^2$ and squares for $(\Delta y)^2$.
Dashed lines are guides to the eye.
Error bars are less than the symbol size.
}
\label{f5}
\end{figure}

To analyze the H displacements parallel to the layer plane
we consider its mean-square displacement (MSD) in 
the $x$ and $y$ directions.
In Fig.~5 we show the temperature dependence of these MSDs,
as derived from MD simulations. 
In our coordinate system, the projection of the C--H bond 
on the layer plane is parallel to the $y$ axis.
This means that the $y$ axis corresponds to the direction 
from the C bound to H to the center of the vacancy (missing C atom),
which coincides with the so-called armchair direction in the
graphene sheet. 
Then, the MSD in the $x$ direction (parallel to the zigzag
direction of graphene) is related to bending of 
the C--H bond (displacement perpendicular to the C--H axis). 
Within the precision of our results, we find that $(\Delta x)^2$ 
increases linearly as the temperature is raised.

For the MSD in the $y$ direction, however, we observe a clear
deviation from linearity. 
The main reason for this behavior of $(\Delta y)^2$ is the
onset of H jumps from one side to the other of the graphene
layer at $T >$ 400~K in the course of the MD simulations.
This means that during the jump process both bonded H and C
atoms cross the plane $z = 0$, where their $y$ coordinates 
change with respect to those corresponding to the steady
configurations at one or the other side (see Sec.~V).
The jump process does not appreciably affect the mean $x$ 
coordinate of the linked H and C atoms.
The change in the mean $y$ coordinate of H, 
$\langle y_{\rm H} \rangle$, amounts to 0.04 \AA, which 
although not very large is enough to cause a superlinear
increase in $(\Delta y)^2$ vs the temperature.
In fact, we obtain in the region from 300 to 1200~K a
dependence $(\Delta y)^2 \sim T^{\alpha}$ with an exponent
$\alpha = 1.24$.

We have calculated the vibrational DOS for H bound to a C atom 
close to a vacancy, using the HLR method described in Sec.~II.C.
The result obtained with this procedure for $T$ = 300~K is shown
in Fig.~2 (continuous line). This DOS for H vibrations was derived 
from atomic displacements around their equilibrium positions 
in MD trajectories including $1.2 \times 10^7$ steps. 
This length of the trajectories was necessary for a precise
determination of the frequencies, with statistical error
bars of less then 5~cm$^{-1}$.

The clearest feature in the vibrational DOS of H is the C--H stretching
mode, which appears at frequencies well above the rest of vibrations
in the defective material. For this stretching vibration we find
at 300~K a frequency of 2990~cm$^{-1}$ (see Fig.~2),
which means an anharmonic redshift of 172~cm$^{-1}$ with respect to
the harmonic calculation presented above in Sec.~II (dashed line).
Using the so-called forced vibrational method, 
Islam {\em et al.}\cite{is14} found a stretching frequency 
of about 2900 cm$^{-1}$ for C--H stretching vibrations in defective 
graphene nanoribbons.
For comparison, we mention that the HLR method has been previously
applied to study vibrational modes in molecular systems. Thus,
for C--H stretching modes in the ethyl radical (C$_2$H$_5$), it was 
found at $T$ = 25~K a redshift in the range from 270 to 
310~cm$^{-1}$.\cite{lo03}
For the DOS of H in defective graphene we observe
in the range between 400 and 1500~cm$^{-1}$ changes in
the position and intensity of the modes respect the harmonic result, 
with the most prominent peaks appearing now at 1005 and 1285~cm$^{-1}$.

\begin{figure}
\vspace{-0.5cm}
\includegraphics[width=7cm]{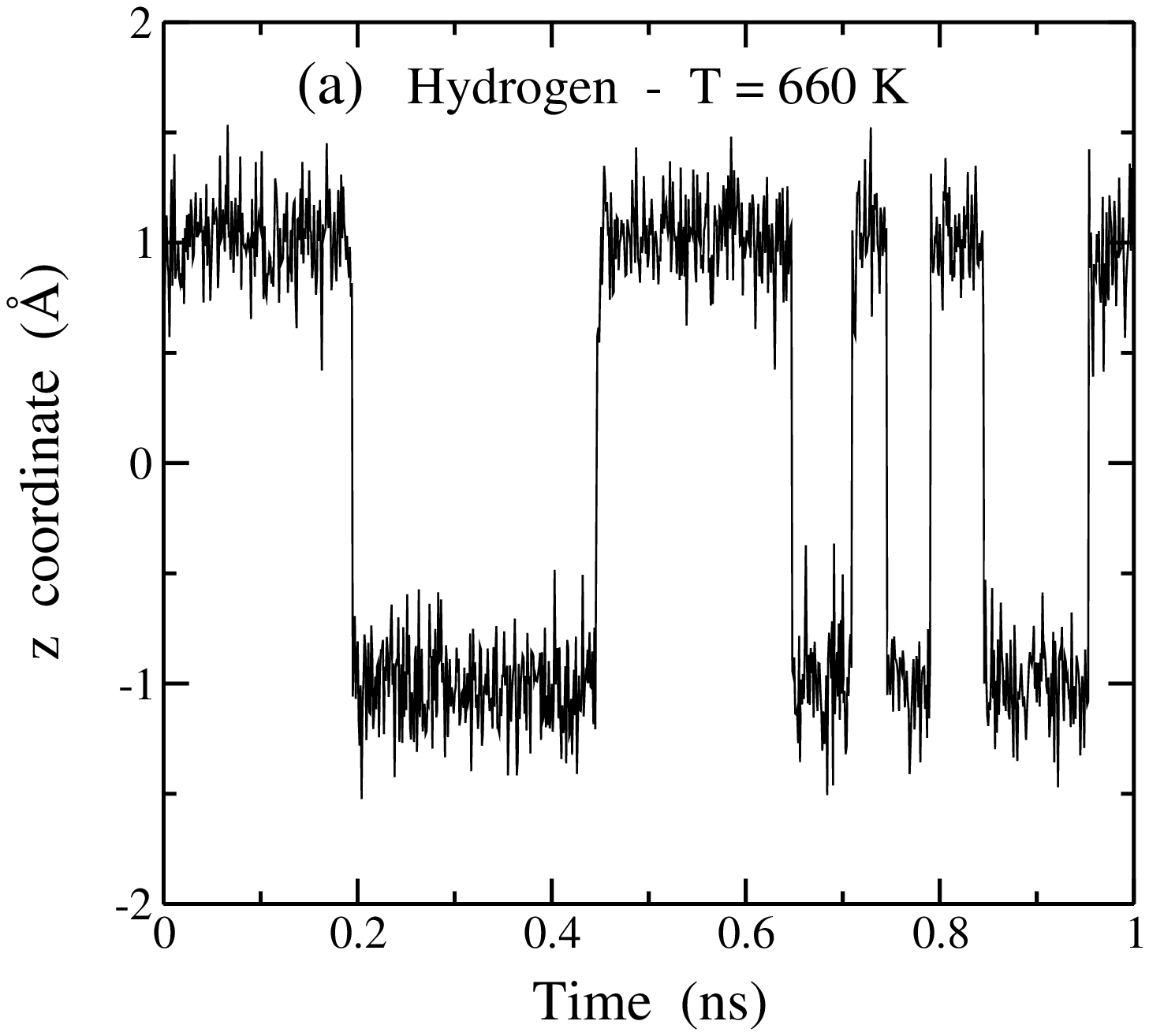}
\includegraphics[width=7cm]{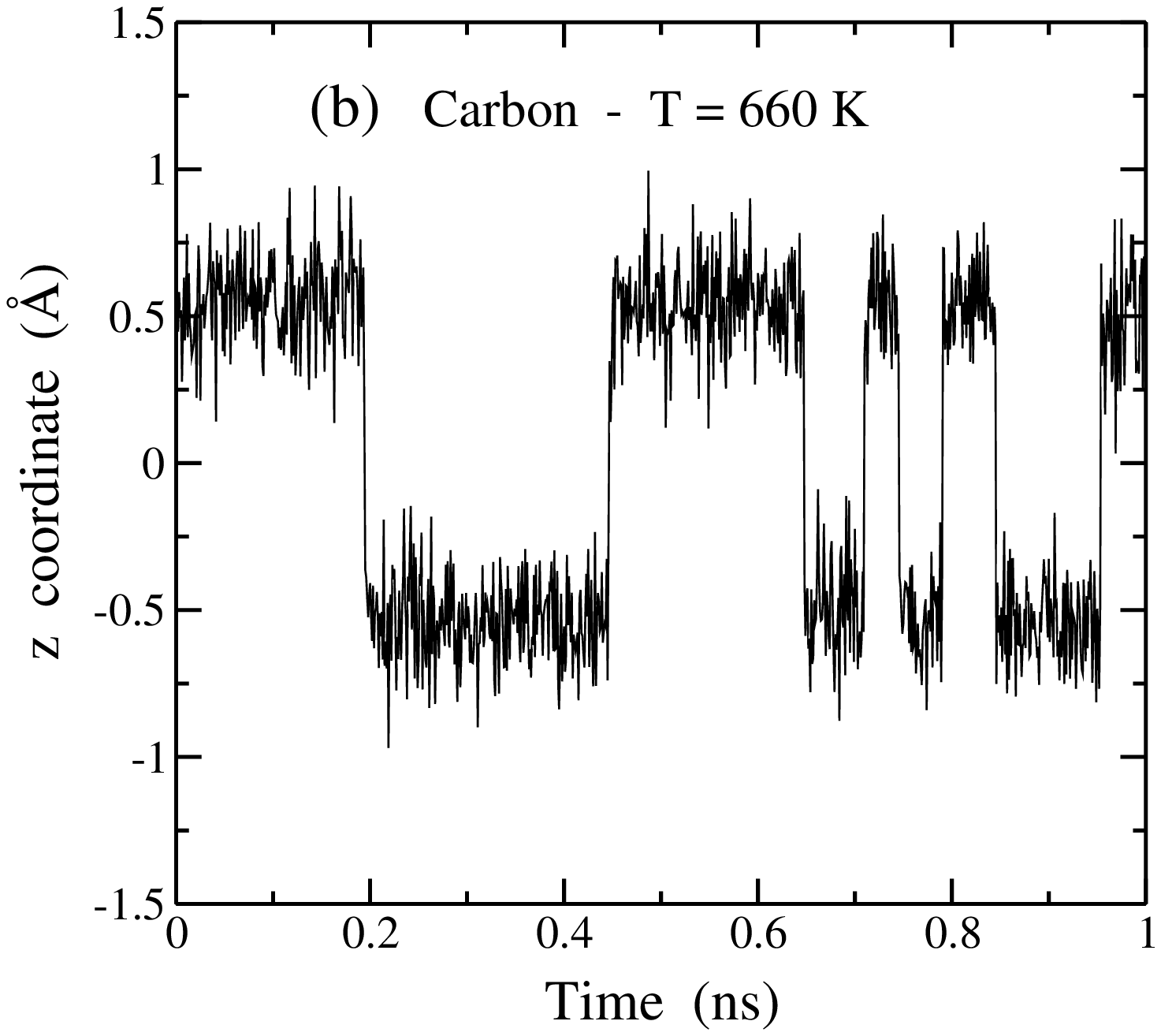}
\vspace{-0.5cm}
\caption{
Coordinate $z$ of (a) hydrogen and (b) carbon along a MD simulation
run at 660 K. Both atoms form a C--H bond.
The data shown include $2 \times 10^6$ MD steps, corresponding to
a time interval of 1 ns.
}
\label{f6}
\end{figure}

\section{Hydrogen dynamics}

We now present results of MD simulations for motion of one 
hydrogen atom in defective graphene at finite temperatures. 
From the calculations at zero temperature, 
we can expect jumps of H from one side to the opposite one
of the graphene layer.
In Fig.~6(a) we present the evolution of the $z$ coordinate of
hydrogen, $z_{\rm H}$, along a MD simulation run at $T = 660$~K. 
The time interval displayed here amounts to 1 ns, i.e., 
$2 \times 10^6$ MD steps.
Even though $z_{\rm H}$ appreciably fluctuates due to
thermal motion of H, we clearly observe that the hydrogen
atom resides most of the time around $z_{\rm H}$ = 1 or 
--1 \AA. Around both plateaus we find fluctuations of
$z_{\rm H}$ with a MSD $(\Delta z_{\rm H})^2$ = 0.04~\AA$^2$.

The evolution of the $z$ coordinate of the C atom bound to H,
$z_{\rm C}$, is displayed in Fig~6(b) for the same time interval 
as in Fig~6(a) for H. Jumps of this C atom between the upper and 
lower side of the graphene sheet occur in synchrony with those 
of H, keeping a C--H distance of 1.11~\AA.
At 660~K, we find $\langle |z_{\rm C}| \rangle$ = 0.53~\AA.

At $T \geq 500$~K we have observed up-down jumps of H and C atoms 
along the MD simulations similar to those presented in Fig.~6.
At lower $T$, we have not found any such event along our simulations,
with a duration of 10 ns. This means that at room temperature 
($T =$ 300~K) one expects a jump rate $\nu \lesssim$ 1/(10 ns) = 
$10^8$ s$^{-1}$. 
At $T \geq 500$~K the jump frequency $\nu$ can be reliably estimated
from the number of up-down jumps observed along the simulations.

\begin{figure}
\vspace{-0.5cm}
\includegraphics[width=7cm]{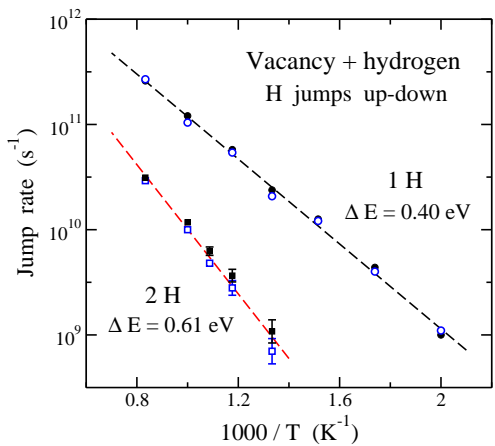}
\vspace{-0.5cm}
\caption{
Up-down jump rate $\nu$ for H hopping between $z_{\rm H} > 0$
and $z_{\rm H} < 0$ vs the inverse temperature for hydrogen
close to a vacancy in graphene.
Circles and squares represent the jump rate obtained for
one H and two H atoms in a C vacancy, respectively.
In both cases, solid symbols indicate data obtained from the
statistics of jumps observed along the MD trajectories.
Open symbols represent results for $\nu$ derived from the
autocorrelation function $G(\tau)$.  When not shown,
error bars are in the order of the symbol size.
Dashed lines are least-square fits to the data points.
$\Delta E$ is the effective energy barrier for H jumps
(see text for details).
}
\label{f7}
\end{figure}

For a number of atomic jumps $N_{+-}$ in a simulation time $\Omega$, 
our estimation for the rate is $\nu = N_{+-} / \Omega$.
In this way, we have calculated the rate $\nu$ at several 
temperatures up to 1200~K from the jumps occurring in our MD 
simulations.  In Fig.~7 we present an Arrhenius-type plot of the 
obtained jump rate vs the inverse temperature.
Solid circles represent the frequency $\nu$ obtained from the
observed number of jumps $N_{+-}$ along simulation runs of 
$2 \times 10^7$ MDS ($\Omega$ = 10 ns).
These results can be well fitted to an expression 
$\nu \propto \exp (-\Delta E / k_B T)$,
with an energy barrier $\Delta E$ = 0.40(1) eV.
This effective barrier is close to that found for crossing
the layer plane at $T = 0$, $\Delta E$ = 0.37~eV (see Sec.~III).

An alternative procedure to calculate the jump rate $\nu$ is based
on the autocorrelation function $G(\tau)$ for the coordinate $z_{\rm H}$.
We define this time function as
\begin{equation}
   G(\tau) = \frac{\langle z_{\rm H}(t) z_{\rm H}(t + \tau) \rangle}
         {\langle z_{\rm H}(t)^2 \rangle}  \, ,
\label{gtau}
\end{equation}
where the mean values are taken along a MD simulation run.
Assuming that $z_{\rm H}$ is a stochastic variable taking values $c$ 
and $-c$ with equal probability and a hopping frequency $\nu$, we have
for a given initial condition $z_{\rm H}(0)$ (see Appendix):
\begin{equation}
   \frac{d \langle z_{\rm H}(t) \rangle}{d t} = 
	     - 2 \nu \, \langle z_{\rm H}(t) \rangle   \, ,
\end{equation}
which yields
\begin{equation}
	G(\tau) = \exp(-2 \nu \tau)   \, .
\label{gtau2}
\end{equation}

\begin{figure}
\vspace{-0.5cm}
\includegraphics[width=7cm]{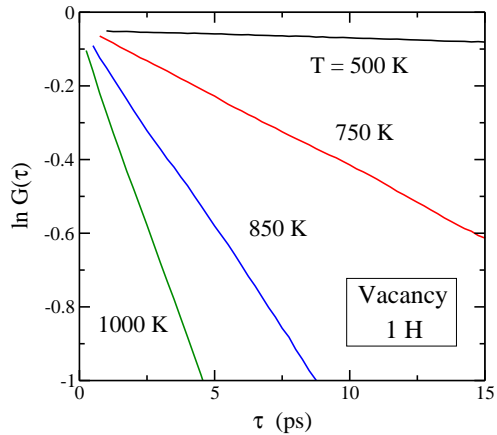}
\vspace{-0.5cm}
\caption{
Autocorrelation function $G(\tau)$ for the $z$ coordinate
of a single H atom in a carbon vacancy at several temperatures,
From top to bottom: $T$ = 500, 750, 850, and 1000 K.
}
\label{f8}
\end{figure}

The autocorrelation function $G(\tau)$ for a single hydrogen atom has 
been calculated from the H trajectories obtained in our MD simulations
at various temperatures, following the definition in Eq.~(\ref{gtau}).
The results are presented in Fig.~8 at four temperatures in 
a logarithmic plot.
From the slope of these lines we obtain the frequency $\nu$ using
Eq.~(\ref{gtau2}).
Note that the lines displayed in Fig.~8 do not extrapolate to
zero for $\tau \to 0$ (i.e., $G(\tau) \to 1$), due to thermal motion
around the energy minima, which causes a fast decrease of $G(\tau)$
at short times to values $G(\tau) \sim 0.9$.
Thus, we observe in fact $G(\tau) = C \exp(-2 \nu \tau)$, 
with $C < 1$.

Open circles in Fig.~7 indicate the rate $\nu$ obtained from 
the autocorrelation function $G(\tau)$ at various temperatures.
The results found with this procedure and those derived from
direct enumeration of the H jumps along the MD trajectories
(solid circles) agree well, and yield the same effective
energy barrier $\Delta E$ = 0.40(1)~eV for this process.
We have thus derived this effective barrier in two independent
ways, which provides us with a consistency check for our
calculations.
From the data presented in Fig.~7, we can estimate the expected
jump rate at room temperature from extrapolation of the
results at higher temperatures. From the linear fit in
the Arrhenius plot of Fig.~7, we find 
a ratio $\nu = 2 \times 10^6 \, {\rm s}^{-1}$ at $T$ = 300~K.

At this point,
one may ask if the hydrogen jumps are uncorrelated or there
is some correlation between them, in the sense that the
atomic arrangement after a jump ``remembers'' in some way the
configuration before the jump. This could favor a return to
the old configuration.\cite{ph91}
The presence or lack of correlations between H jumps can be 
analyzed by studying the distribution of the number of jumps
for time intervals of a given length $\Pi$.
For uncorrelated stochastic events, their number in a time
interval follows a Poisson distribution.\cite{fe68}
In our case, this means that for uncorrelated H jumps with
jump rate $\nu$ (mean time between events equal to $1 / \nu$), 
the probability $P(n)$ for the number $n$ of jumps in 
a time interval $\Pi$ should be given by\cite{fe68,fe10}
\begin{equation}
   P(n) = {\rm e}^{- \nu \Pi} \, \frac{(\nu \Pi)^n}{n !}   \; .
\label{poisson}
\end{equation}
The mean value for this probability distribution is $\mu = \nu \Pi$,
i.e., the average number of hydrogen jumps in a time interval $\Pi$
is $\nu \Pi$.
Note that for low temperature and small $\nu$, i.e. $\nu \Pi \ll 1$, 
only $n$ = 0 and 1 will have an appreciable probability. 

\begin{figure}
\vspace{-0.5cm}
\includegraphics[width=7cm]{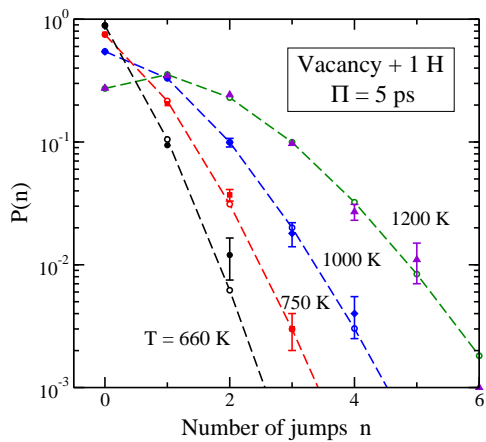}
\vspace{-0.5cm}
\caption{Probability distribution for the number of H jumps in a
time window $\Pi$ = 5 ps. Solid symbols represent results derived
from MD simulations at $T$ = 660 (circles), 750 (squares),
1000 (diamonds), and 1200~K (triangles).
Open circles joined with dashed lines indicate the expected
Poisson distribution corresponding to the same time window $\Pi$.
Error bars of the simulation results, when not shown, are in
the order or less than the symbol size.
}
\label{f9}
\end{figure}

We have split our MD trajectories at several temperatures in time 
intervals of length $\Pi$ = 5 ps ($10^4$ MD steps), and obtained 
frequencies for the number of jumps $n$ in those intervals, from 
where we derived the corresponding probability distribution.
This is shown in Fig.~9, where solid symbols represent results 
yielded by MD simulations at different temperatures.
Open circles and dashed lines connecting them indicate the 
Poisson distribution calculated according 
to Eq.(\ref{poisson}) with the same time interval $\Pi$.
The results of MD simulations follow closely the Poisson
distribution, in agreement with the assumption 
that H jumps behave as uncorrelated events.
For other time intervals $\Pi$, we also found results 
compatible with a Poisson distribution.

Hydrogen jumps from one C atom to another C atom close to the
vacancy can be observed in the MD simulations by following
changes in the atomic $x$ and $y$ coordinates.
However, even at the highest temperatures considered here
the number of this kind of jumps is scarce in a time window
in the order of 10 ns, and the statistics necessary to
precisely define a jump rate is poor.
This is consistent with a relatively large energy barrier 
of 1.54~eV mentioned in Sec.~III for this kind of H jumps.

For $T$ lower than room temperature quantum effects are expected
to appear in the hydrogen dynamics.
In the language of transition-state theory,\cite{fl70} such
quantum effects give rise to a renormalization of the hopping
barrier, which can be effectively lowered in comparison to
the classical result.\cite{su80,sc88,gi88,no97b}
This low-temperature barrier could be investigated by
using transition-state theory along with quantum
path-integral simulations. Some work in this line has been
performed for a hydrogen impurity on a pristine graphene 
sheet\cite{he09a}
as well as in bulk diamond\cite{he07} and silicon.\cite{he97}
This question lies however outside the scope of the present paper.

\section{Two hydrogen atoms in a carbon vacancy}

We have also carried out MD simulations for two H atoms bound 
to two different C atoms close to a vacancy in graphene. In this
case, two dangling bonds are passivated by hydrogen.
According to the calculations at $T = 0$ presented in Sec.~III,
the minimum-energy state corresponds to the H atoms (which we will
label 1 and 2) located at opposite sides of the graphene layer,
each one at a distance of 0.75~\AA\ to the layer plane.

\begin{figure}
\vspace{-0.5cm}
\includegraphics[width=7cm]{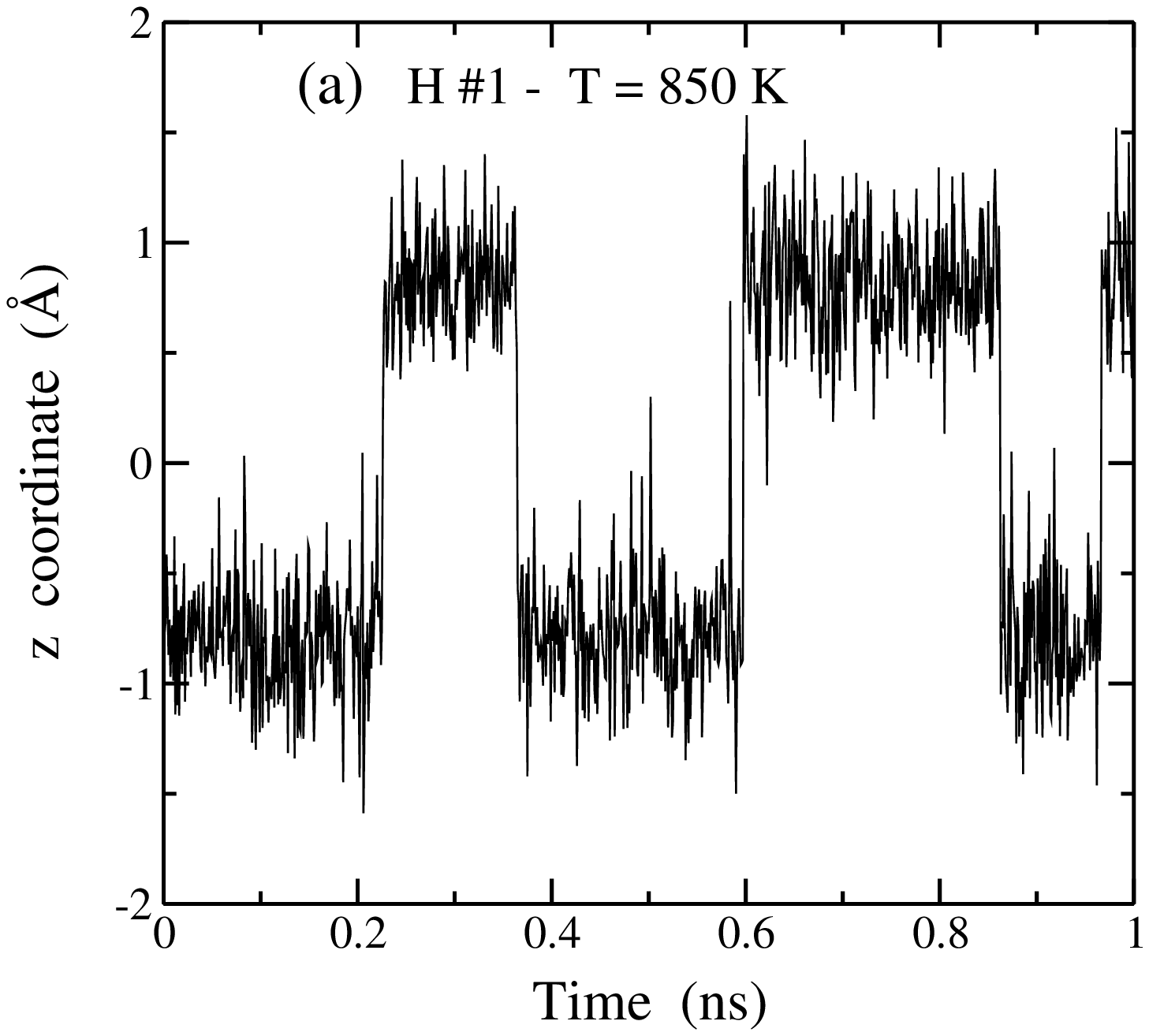}
\includegraphics[width=7cm]{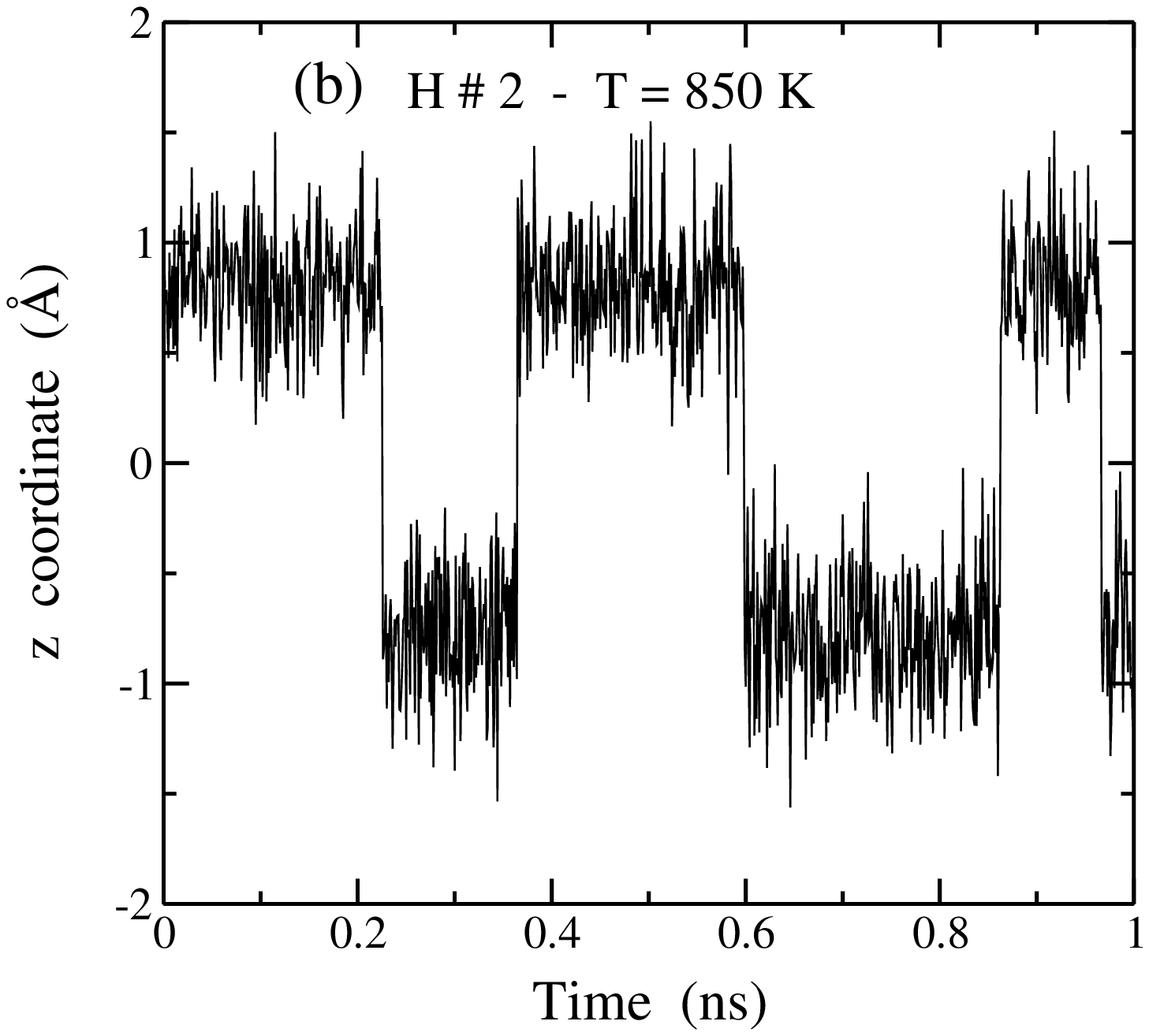}
\vspace{-0.5cm}
\caption{
Coordinate $z_{\rm H}$ of both H atoms close to a C vacancy along
a MD simulation run at $T$ = 850 K.
The data shown in (a) and (b) correspond to H \#1 and H \#2
in a time interval of 1 ns ($2 \times 10^6$ MD steps).
}
\label{f10}
\end{figure}

As in the case of a single H at the vacancy, for two H atoms 
one expects appreciable motion of H at finite temperatures, 
and in particular crossing of the graphene plane.
In Fig.~10 we show the coordinates $z_{\rm H}^{(1)}$ and
$z_{\rm H}^{(2)}$ of both hydrogen atoms 
at $T$ = 850~K, along a MD trajectory in a time interval 
of 1~ns ($2 \times 10^6$ MD steps).
Panels (a) and (b) correspond to hydrogen atoms 1 and 2,
respectively. One observes the correlation in the jumps of
both atoms: they are located at opposite sides of the graphene
layer, and simultaneously (as seen at the scale of the plot) 
cross the layer plane interchanging their coordinate $z_{\rm H}$.
At this temperature, the mean distance of the H atoms to
the plane is $\langle |z_{\rm H}| \rangle$ = 0.81~\AA,
somewhat higher than in the minimum-energy configuration,
with a MSD $(\Delta z_{\rm H})^2$ = 0.07~\AA$^2$.

We have calculated the jump rate of the H atoms using the
same procedures as in the case of a single hydrogen.
The results are presented in Fig.~7 along with the data
corresponding to a single H, discussed above in Sec.~V.
For two H atoms, solid squares correspond to the mean jump
rate $\nu$ obtained from the statistics of jumps observed
along the simulations ($\nu = N_{+-} / \Omega$),
whereas open squares were derived from the correlation 
function $G(\tau)$ for each H atom.
The results of both methods are close one to the other,
and we observe that the solid symbols lie somewhat above 
the open ones, especially at the lowest temperatures considered
here.  This does not seem to be fortuitous, and may be due 
(apart from the poorer statistics at lower temperature) to 
additional correlations in the jump process, not taken into account
in the statistics used in our procedures.
In any case, the whole ensemble of data can be well fitted
to an expression of the form $\nu \propto \exp(- \Delta E / k_B T)$,
with an activation barrier $\Delta E$ = 0.61(3) eV.
This effective energy barrier is close to that obtained at
$T = 0$ of 0.57~eV (see Sec.~III).

Extrapolation to room temperature ($T = 300$~K) of the linear fit 
in the Arrhenius 
plot for the case of two H atoms gives a relatively low jump rate 
$\nu = 7 \times 10^2 {\rm s}^{-1}$.  It is a factor of about
$3 \times 10^3$ less than the jump rate of a single H in a carbon
vacancy at this temperature.

\begin{figure}
\vspace{-0.5cm}
\includegraphics[width=7cm]{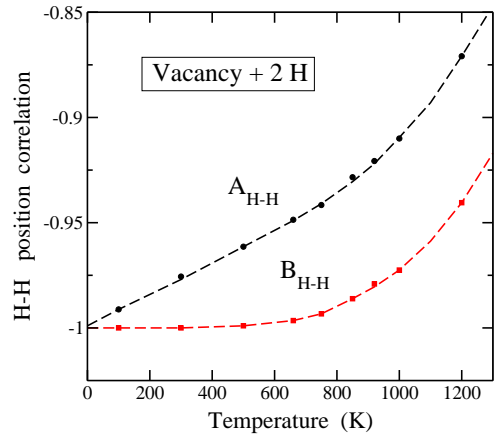}
\vspace{-0.5cm}
\caption{
Temperature dependence of the correlation functions
$A_{\rm H-H}$ and $B_{\rm H-H}$ for the
$z$ coordinate of both H atoms close to a carbon vacancy.
These functions are defined in Eqs.~(\ref{ahh}) and (\ref{bhh}).
Symbols are data points derived from MD simulations.
Dashed lines are guides to the eye.
}
\label{f11}
\end{figure}

To further analyze correlations in the positions of both H atoms
close to a carbon vacancy, and their dependence on temperature, 
we have employed two functions of the coordinates 
$z_{\rm H}^{(1)}$ and $z_{\rm H}^{(2)}$.
The first function, $A_{\rm H-H}$, is defined as: 
\begin{equation}
 A_{\rm H-H} = \frac{ \left< z_{\rm H}^{(1)} z_{\rm H}^{(2)} \right> }
    { \left< |z_{\rm H}^{(1)}| \right>  
       \left< |z_{\rm H}^{(2)}| \right> }    \;  ,
\label{ahh}
\end{equation}
where the brackets indicate averages over the MD trajectories.
In Fig.~11 we present the temperature dependence of $A_{\rm H-H}$
(solid circles), as derived from our simulations.
For the ideal disposition of H atoms fixed on their
minimum-energy positions (classical $T = 0$ limit), one has
$A_{\rm H-H} = -1$. At finite temperatures, 
thermal motion causes an increase in $A_{\rm H-H}$ (a reduction 
of its absolute value), as shown in Fig.~11.
$A_{\rm H-H}$ grows linearly with temperature up to about
800~K, and increases faster at higher $T$.
The main reason for this increase in $A_{\rm H-H}$
is the rise of the numerator in Eq.~(\ref{ahh}),
since thermal fluctuations in the atomic positions
cause uncorrelated motion of both H atoms around their
equilibrium positions.

The second (apparently simpler) function considered here to
analyze correlations in the $z$ coordinates of both
H atoms is defined as
\begin{equation}
 B_{\rm H-H} =  \left<  \frac{ z_{\rm H}^{(1)} } {|z_{\rm H}^{(1)}| }
      \;   \frac{ z_{\rm H}^{(2)} } {|z_{\rm H}^{(2)}| }  \right>  =
      \left< {\rm sgn} (z_{\rm H}^{(1)}) \; {\rm sgn} (z_{\rm H}^{(2)})
      \right> \; ,
\label{bhh}
\end{equation}
where ``sgn'' denotes the sign function, i.e., sgn $z$ = 1 or -1 for
$z > 0$ and $z < 0$, respectively.
Results for $B_{\rm H-H}$ derived from our MD simulations are 
shown in Fig.~11 as solid squares.

The product of both terms inside the brackets in Eq.~(\ref{bhh})
takes values 1 or --1, the former when both H atoms are at the
same side of the graphene layer and the latter when they are
at opposite sites of the layer plane.
Thus, at $T < 700$~K the probability for having simultaneously
both H atoms at the same side is very low, and
$B_{\rm H-H} \approx -1$. At higher temperatures, this
probability grows, mainly due to thermal fluctuations of
the atomic position during the up-down jumps.
Looking together at both functions in Fig.~11, $A_{\rm H-H}$ and 
$B_{\rm H-H}$, it turns out that the former is more sensitive to 
thermal motion than the latter, especially at temperatures lower 
than 800~K, where $B_{\rm H-H}$ is nearly constant and 
$A_{\rm H-H}$ grows linearly as the temperature is raised.

The function $B_{\rm H-H}$ is directly related to the probability
for finding both H atoms on the same side of the graphene layer.
Calling $Q$ this probability, one has
\begin{equation}
    Q = \frac12  \left( 1 + B_{\rm H-H} \right)  \; .
\end{equation}
In the low-$T$ classical limit one has $Q = 0$.
At $T$ = 1000~K, we obtain $B_{\rm H-H} = -0.972$ and
$Q = 0.014$, still a rather small probability at this high 
temperature.

We note that at low $T$, the treatment of H atoms as
distinguishable entities is no longer possible, as they
have to be considered as quantum identical atoms,
and classical trajectories for both of them are not
realistic. At $T \gtrsim 300$~K, however,
the quantum exchange probability is very low, so that
such trajectories are a reliable description of this 
system.  A classical treatment at these temperatures is 
favored by the fact that the defect complexes that really
hop include not only hydrogen atoms, but also the
heavier carbon atoms bound to them.

\section{Summary}

We have studied the dynamics of atomic hydrogen in defective
graphene. In particular, we considered crossing of a graphene 
layer by H atoms in the presence of carbon vacancies.
This provides us with essential information on a relevant part of
the dynamical procces leading to hydrogen permeation in graphene.

It has been shown that
TB MD simulations of hydrogen in defective graphene constitute
a reliable tool to study the atomic motion
in a wide range of temperatures, employing an interatomic potential
fitted to {\em ab-initio} calculations.
As a consequence of the large relaxation of the nearest C atoms, 
the hydrogen dynamics in graphene requires a concomitant motion 
of these atoms.
Thus, H jumps from one side of the layer plane to the other
are in fact a cooperative process engaging the impurity and 
the nearest host atoms.

For an H atom near a vacancy, the minimum-energy configuration
corresponds to the impurity located off-plane at a distance
of 0.77~\AA\ from the graphene layer. The C atom attached
to H also relaxes off-plane to $z_{\rm C}$ = 0.40~\AA. 
The C--H bond forms an angle of 19.5 degrees with the 
graphene plane.
Our results at $T = 0$ derived from the TB Hamiltonian display 
good agreement with those obtained from {\em ab-initio} DFT
calculations.

At finite temperatures, the coordinate $z_{\rm H}$ presents 
plateaus along trajectories generated by MD simulations, with 
jumps from one plateau to the other corresponding to hydrogen
crossing the layer plane 
(interchange between $z_{\rm H} > 0$ and $z_{\rm H} < 0$).
Jump rates at different temperatures were obtained using two
methods based on the jump statistics along the MD trajectories 
and the autocorrelation function $G(\tau)$. Both procedures give
results consistent with one another.
Analysis of the impurity jump rate as a function of the
inverse temperature allowed us to calculate an effective energy
barrier for traversing the layer plane at finite temperatures.
For an H atom near a vacancy, we found a barrier 
of 0.40~eV, with a jump frequency 
$\nu = 2 \times 10^6$~s$^{-1}$ at room temperature.

For two H atoms, we observe a concerted motion with concurrent 
passage through the vacancy.
In this case, the effective barrier is found to be 0.61~eV, 
with $\nu = 7 \times 10^2$~s$^{-1}$ at 300~K.
The C atoms linked to H atoms equally move from one to the other
side of the graphene layer, thus preserving the C--H bonds along
this dynamic process.

In this paper, we have focused on the crossing mechanism for atomic 
hydrogen.  The whole process of hydrogen permeation through carbon 
vacancies in graphene would include also H$_2$ dissociation and 
atomic diffusion on the surface, prior to the layer crossing, 
as well as subsequent recombination and desorption.

An interesting extension of this work could be the consideration
of quantum effects in the hydrogen dynamics. In particular,
application of transition-state theory, based on Feynman
path integrals may be employed to study the renormalization
of the classical energy barriers and the jump rates at $T$ lower
than room temperature. \\  \\

\noindent
{\bf CRediT author contribution statement}  \\

Carlos P. Herrero: Data curation, Investigation, Validation, Original draft

Jos\'e A. Verg\'es: Methodology, Investigation, Validation

Rafael Ram\'irez: Methodology, Software, Investigation, Validation  \\

\noindent
{\bf Declaration of Competing Interest}  \\

The authors declare that they have no known competing financial
interests or personal relationships that could have appeared to
influence the work reported in this paper.  \\

\begin{acknowledgments}
This work was supported by Ministerio de Ciencia e Innovaci\'on
(Spain) through Grant PGC2018-096955-B-C44.
\end{acknowledgments}

%  -----------------------------------------------------------------

\appendix

\section{Autocorrelation function}

The autocorrelation function for the coordinate $z_{\rm H}$ of 
hydrogen is defined as 
\begin{equation}
 G(\tau) = \frac{\langle z_{\rm H}(t) z_{\rm H}(t + \tau) \rangle}
      {\langle z_{\rm H}(t)^2 \rangle}
\end{equation}
with
\begin{equation}
 \langle z_{\rm H}(t) z_{\rm H}(t + \tau) \rangle =
 \frac{1}{\Omega}  \int_0^{\Omega} z_{\rm H}(t) z_{\rm H}(t + \tau) dt  \; .
\end{equation}

To derive an analytical expression for the autocorrelation function,
we consider a random variable $z_{\rm H}$ taking values $c$ and $-c$ with
equal probability, and a time sequence controlled by a
stochastic process with a rate $\nu$ for jumps between $c$ and $-c$.
We will call $P_+(t)$ and $P_-(t)$ the probabilities of having
$z_{\rm H} = c$ or $-c$ at time $t$, respectively, 
with $P_+(t) + P_-(t) = 1$.  
Given an initial condition $z_{\rm H}(0)$, we have for time $t$:
\begin{equation}
     \langle z_{\rm H}(t) \rangle = 
      c P_+(t) - c P_-(t) =  c \left[ 2 P_+(t) - 1 \right]   \; ,
\end{equation}
and
\begin{equation}
	\frac{d \langle z_{\rm H}(t) \rangle}{d t} =
      2 c \, \frac{d P_+(t)}{d t}  \; .
\label{dzh}
\end{equation}
Here $\langle z_{\rm H}(t) \rangle$ indicates the mean value of
$z_{\rm H}$ at time $t$ for all stochastic trajectories starting
at $z_{\rm H}(0)$.

For a time interval $\Delta t$, the mean number of jumps is
$\nu \Delta t$. Then, for small $\Delta t$ (i.e., $\nu \Delta t \ll 1$) 
the probability $P_+(t + \Delta t)$ can be written as
\begin{equation}
   P_+(t + \Delta t) = P_+(t) + \left[ P_-(t) - P_+(t) \right] 
	  \nu \Delta t  \, ,
\end{equation}
and a similar expression holds for $P_-(t + \Delta t)$.
This means that the time derivative of $P_+(t)$ is given by
\begin{equation}
 \frac{d P_+(t) }{dt} = \left[ P_-(t) - P_+(t) \right] \nu  \; ,
\end{equation}
or
\begin{equation}
   \frac{d P_+(t)}{dt} = \left[ 1 - 2 P_+(t) \right] \nu  \; .
\label{dp+}
\end{equation}
Then, from Eqs.~(\ref{dzh}) and (\ref{dp+}) one finds
\begin{equation}
 \frac{d \langle z_{\rm H}(t) \rangle}{d t} = 
	- 2 \nu \, \langle z_{\rm H}(t) \rangle
\end{equation}
and
\begin{equation}
   \langle z_{\rm H}(t) \rangle = z_{\rm H}(0) \exp(-2 \nu t)  \; .
\end{equation}
Note that for long $t$ ($\nu t \gg 1$), both $P_+$ and $P_-$
converge to 0.5, and $\langle z_{\rm H}(t) \rangle \to 0$.

Finally, we find for the autocorrelation function
\begin{equation}
 G(\tau) = \frac{\langle z_{\rm H}(t) z_{\rm H}(t + \tau) \rangle}
           {\langle z_{\rm H}(t)^2 \rangle}  =
           \frac{\langle z_{\rm H}(0) z_{\rm H}(\tau) \rangle} 
           {\langle z_{\rm H}(0)^2 \rangle} = \exp(-2 \nu \tau)  \; ,
\end{equation}
where we have used the time translation invariance for the origin
of the stochastic trajectories.

%  -------------------------------------------------------------------


\begin{thebibliography}{89}
\expandafter\ifx\csname natexlab\endcsname\relax\def\natexlab#1{#1}\fi
\expandafter\ifx\csname bibnamefont\endcsname\relax
  \def\bibnamefont#1{#1}\fi
\expandafter\ifx\csname bibfnamefont\endcsname\relax
  \def\bibfnamefont#1{#1}\fi
\expandafter\ifx\csname citenamefont\endcsname\relax
  \def\citenamefont#1{#1}\fi
\expandafter\ifx\csname url\endcsname\relax
  \def\url#1{\texttt{#1}}\fi
\expandafter\ifx\csname urlprefix\endcsname\relax\def\urlprefix{URL }\fi
\providecommand{\bibinfo}[2]{#2}
\providecommand{\eprint}[2][]{\url{#2}}

\bibitem[{\citenamefont{Pearton et~al.}(1992)\citenamefont{Pearton, Corbett,
  and Stavola}}]{pe92}
\bibinfo{author}{\bibfnamefont{S.~J.} \bibnamefont{Pearton}},
  \bibinfo{author}{\bibfnamefont{J.~W.} \bibnamefont{Corbett}},
  \bibnamefont{and} \bibinfo{author}{\bibfnamefont{M.}~\bibnamefont{Stavola}},
  \emph{\bibinfo{title}{Hydrogen in Crystalline Semiconductors}}
  (\bibinfo{publisher}{Springer}, \bibinfo{address}{Berlin},
  \bibinfo{year}{1992}).

\bibitem[{\citenamefont{Estreicher}(1995)}]{es95}
\bibinfo{author}{\bibfnamefont{S.~K.} \bibnamefont{Estreicher}},
  \bibinfo{journal}{Mater. Sci. Eng.} \textbf{\bibinfo{volume}{R14}},
  \bibinfo{pages}{319} (\bibinfo{year}{1995}).

\bibitem[{\citenamefont{Whitener}(2018)}]{wh18}
\bibinfo{author}{\bibfnamefont{K.~E.} \bibnamefont{Whitener},
  \bibfnamefont{Jr.}}, \bibinfo{journal}{J. Vac. Sci. Technol. A}
  \textbf{\bibinfo{volume}{36}}, \bibinfo{pages}{05G401}
  (\bibinfo{year}{2018}).

\bibitem[{\citenamefont{Fan et~al.}(2015)\citenamefont{Fan, Zhang, and
  Zhang}}]{fa15}
\bibinfo{author}{\bibfnamefont{X.}~\bibnamefont{Fan}},
  \bibinfo{author}{\bibfnamefont{G.}~\bibnamefont{Zhang}}, \bibnamefont{and}
  \bibinfo{author}{\bibfnamefont{F.}~\bibnamefont{Zhang}},
  \bibinfo{journal}{Chem. Soc. Rev.} \textbf{\bibinfo{volume}{44}},
  \bibinfo{pages}{3023} (\bibinfo{year}{2015}).

\bibitem[{\citenamefont{Hu et~al.}(2017)\citenamefont{Hu, Yao, and
  Wang}}]{hu17}
\bibinfo{author}{\bibfnamefont{M.}~\bibnamefont{Hu}},
  \bibinfo{author}{\bibfnamefont{Z.}~\bibnamefont{Yao}}, \bibnamefont{and}
  \bibinfo{author}{\bibfnamefont{X.}~\bibnamefont{Wang}},
  \bibinfo{journal}{Ind. Eng. Chem. Res.} \textbf{\bibinfo{volume}{56}},
  \bibinfo{pages}{3477} (\bibinfo{year}{2017}).

\bibitem[{\citenamefont{Dillon and Heben}(2001)}]{di01}
\bibinfo{author}{\bibfnamefont{A.~C.} \bibnamefont{Dillon}} \bibnamefont{and}
  \bibinfo{author}{\bibfnamefont{M.~J.} \bibnamefont{Heben}},
  \bibinfo{journal}{Appl. Phys. A} \textbf{\bibinfo{volume}{72}},
  \bibinfo{pages}{133} (\bibinfo{year}{2001}).

\bibitem[{\citenamefont{Tozzini and Pellegrini}(2013)}]{to13}
\bibinfo{author}{\bibfnamefont{V.}~\bibnamefont{Tozzini}} \bibnamefont{and}
  \bibinfo{author}{\bibfnamefont{V.}~\bibnamefont{Pellegrini}},
  \bibinfo{journal}{Phys. Chem. Chem. Phys.} \textbf{\bibinfo{volume}{15}},
  \bibinfo{pages}{80} (\bibinfo{year}{2013}).

\bibitem[{\citenamefont{Kag et~al.}(2021)\citenamefont{Kag, Luhadiya, Patil,
  and Kundalwal}}]{ka21}
\bibinfo{author}{\bibfnamefont{D.}~\bibnamefont{Kag}},
  \bibinfo{author}{\bibfnamefont{N.}~\bibnamefont{Luhadiya}},
  \bibinfo{author}{\bibfnamefont{N.~D.} \bibnamefont{Patil}}, \bibnamefont{and}
  \bibinfo{author}{\bibfnamefont{S.~I.} \bibnamefont{Kundalwal}},
  \bibinfo{journal}{Int. J. Hydrogen Energy} \textbf{\bibinfo{volume}{46}},
  \bibinfo{pages}{22599} (\bibinfo{year}{2021}).

\bibitem[{\citenamefont{Sunnardianto et~al.}(2021)\citenamefont{Sunnardianto,
  Bokas, Hussein, Walters, Moultos, and Dey}}]{su21}
\bibinfo{author}{\bibfnamefont{G.~K.} \bibnamefont{Sunnardianto}},
  \bibinfo{author}{\bibfnamefont{G.}~\bibnamefont{Bokas}},
  \bibinfo{author}{\bibfnamefont{A.}~\bibnamefont{Hussein}},
  \bibinfo{author}{\bibfnamefont{C.}~\bibnamefont{Walters}},
  \bibinfo{author}{\bibfnamefont{O.~A.} \bibnamefont{Moultos}},
  \bibnamefont{and} \bibinfo{author}{\bibfnamefont{P.}~\bibnamefont{Dey}},
  \bibinfo{journal}{Int. J. Hydrogen Energy} \textbf{\bibinfo{volume}{46}},
  \bibinfo{pages}{5485} (\bibinfo{year}{2021}).

\bibitem[{\citenamefont{Miao et~al.}(2013)\citenamefont{Miao, Nardelli, Wang,
  and Liu}}]{mi13}
\bibinfo{author}{\bibfnamefont{M.}~\bibnamefont{Miao}},
  \bibinfo{author}{\bibfnamefont{M.~B.} \bibnamefont{Nardelli}},
  \bibinfo{author}{\bibfnamefont{Q.}~\bibnamefont{Wang}}, \bibnamefont{and}
  \bibinfo{author}{\bibfnamefont{Y.}~\bibnamefont{Liu}},
  \bibinfo{journal}{Phys. Chem. Chem. Phys.} \textbf{\bibinfo{volume}{15}},
  \bibinfo{pages}{16132} (\bibinfo{year}{2013}).

\bibitem[{\citenamefont{Tsetseris and Pantelides}(2014)}]{ts14}
\bibinfo{author}{\bibfnamefont{L.}~\bibnamefont{Tsetseris}} \bibnamefont{and}
  \bibinfo{author}{\bibfnamefont{S.~T.} \bibnamefont{Pantelides}},
  \bibinfo{journal}{Carbon} \textbf{\bibinfo{volume}{67}}, \bibinfo{pages}{58}
  (\bibinfo{year}{2014}).

\bibitem[{\citenamefont{Sun et~al.}(2020)\citenamefont{Sun, Yang, Kuang,
  Stebunov, Xiong, Yu, Nair, Katsnelson, Yuan, Grigorieva et~al.}}]{su20b}
\bibinfo{author}{\bibfnamefont{P.~Z.} \bibnamefont{Sun}},
  \bibinfo{author}{\bibfnamefont{Q.}~\bibnamefont{Yang}},
  \bibinfo{author}{\bibfnamefont{W.~J.} \bibnamefont{Kuang}},
  \bibinfo{author}{\bibfnamefont{Y.~V.} \bibnamefont{Stebunov}},
  \bibinfo{author}{\bibfnamefont{W.~Q.} \bibnamefont{Xiong}},
  \bibinfo{author}{\bibfnamefont{J.}~\bibnamefont{Yu}},
  \bibinfo{author}{\bibfnamefont{R.~R.} \bibnamefont{Nair}},
  \bibinfo{author}{\bibfnamefont{M.~I.} \bibnamefont{Katsnelson}},
  \bibinfo{author}{\bibfnamefont{S.~J.} \bibnamefont{Yuan}},
  \bibinfo{author}{\bibfnamefont{I.~V.} \bibnamefont{Grigorieva}},
  \bibnamefont{et~al.}, \bibinfo{journal}{Nature}
  \textbf{\bibinfo{volume}{579}}, \bibinfo{pages}{229} (\bibinfo{year}{2020}).

\bibitem[{\citenamefont{Gupta et~al.}(2018)\citenamefont{Gupta, Kumar, and
  Ray}}]{gu18b}
\bibinfo{author}{\bibfnamefont{V.}~\bibnamefont{Gupta}},
  \bibinfo{author}{\bibfnamefont{A.}~\bibnamefont{Kumar}}, \bibnamefont{and}
  \bibinfo{author}{\bibfnamefont{N.}~\bibnamefont{Ray}},
  \bibinfo{journal}{Pramana J. Phys.} \textbf{\bibinfo{volume}{91}},
  \bibinfo{pages}{64} (\bibinfo{year}{2018}).

\bibitem[{\citenamefont{Kroes et~al.}(2017)\citenamefont{Kroes, Fasolino, and
  Katsnelson}}]{kr17}
\bibinfo{author}{\bibfnamefont{J.~M.~H.} \bibnamefont{Kroes}},
  \bibinfo{author}{\bibfnamefont{A.}~\bibnamefont{Fasolino}}, \bibnamefont{and}
  \bibinfo{author}{\bibfnamefont{M.~I.} \bibnamefont{Katsnelson}},
  \bibinfo{journal}{Phys. Chem. Chem. Phys.} \textbf{\bibinfo{volume}{19}},
  \bibinfo{pages}{5813} (\bibinfo{year}{2017}).

\bibitem[{\citenamefont{Poltavsky et~al.}(2018)\citenamefont{Poltavsky, Zheng,
  Mortazavi, and Tkatchenko}}]{po18b}
\bibinfo{author}{\bibfnamefont{I.}~\bibnamefont{Poltavsky}},
  \bibinfo{author}{\bibfnamefont{L.}~\bibnamefont{Zheng}},
  \bibinfo{author}{\bibfnamefont{M.}~\bibnamefont{Mortazavi}},
  \bibnamefont{and}
  \bibinfo{author}{\bibfnamefont{A.}~\bibnamefont{Tkatchenko}},
  \bibinfo{journal}{J. Chem. Phys.} \textbf{\bibinfo{volume}{148}},
  \bibinfo{pages}{204707} (\bibinfo{year}{2018}).

\bibitem[{\citenamefont{Mazzuca and Haut}(2018)}]{ma18b}
\bibinfo{author}{\bibfnamefont{J.~W.} \bibnamefont{Mazzuca}} \bibnamefont{and}
  \bibinfo{author}{\bibfnamefont{N.~K.} \bibnamefont{Haut}},
  \bibinfo{journal}{J. Chem. Phys.} \textbf{\bibinfo{volume}{148}},
  \bibinfo{pages}{224301} (\bibinfo{year}{2018}).

\bibitem[{\citenamefont{Bartolomei et~al.}(2019)\citenamefont{Bartolomei,
  Hernandez, Campos-Martinez, and Hernandez-Lamoneda}}]{ba19}
\bibinfo{author}{\bibfnamefont{M.}~\bibnamefont{Bartolomei}},
  \bibinfo{author}{\bibfnamefont{M.}~\bibnamefont{Hernandez},
  \bibfnamefont{I}},
  \bibinfo{author}{\bibfnamefont{J.}~\bibnamefont{Campos-Martinez}},
  \bibnamefont{and}
  \bibinfo{author}{\bibfnamefont{R.}~\bibnamefont{Hernandez-Lamoneda}},
  \bibinfo{journal}{Carbon} \textbf{\bibinfo{volume}{144}},
  \bibinfo{pages}{724} (\bibinfo{year}{2019}).

\bibitem[{\citenamefont{Xu et~al.}(2019)\citenamefont{Xu, Jiang, Shen, Li,
  Wang, and Meng}}]{xu19}
\bibinfo{author}{\bibfnamefont{J.}~\bibnamefont{Xu}},
  \bibinfo{author}{\bibfnamefont{H.}~\bibnamefont{Jiang}},
  \bibinfo{author}{\bibfnamefont{Y.}~\bibnamefont{Shen}},
  \bibinfo{author}{\bibfnamefont{X.-Z.} \bibnamefont{Li}},
  \bibinfo{author}{\bibfnamefont{E.~G.} \bibnamefont{Wang}}, \bibnamefont{and}
  \bibinfo{author}{\bibfnamefont{S.}~\bibnamefont{Meng}},
  \bibinfo{journal}{Nature Commun.} \textbf{\bibinfo{volume}{10}},
  \bibinfo{pages}{3971} (\bibinfo{year}{2019}).

\bibitem[{\citenamefont{Griffin et~al.}(2020)\citenamefont{Griffin, Mogg, Hao,
  Kalon, Bacaksiz, Lopez-Polin, Zhou, Guarochico, Cai, Neumann et~al.}}]{gr20}
\bibinfo{author}{\bibfnamefont{E.}~\bibnamefont{Griffin}},
  \bibinfo{author}{\bibfnamefont{L.}~\bibnamefont{Mogg}},
  \bibinfo{author}{\bibfnamefont{G.-P.} \bibnamefont{Hao}},
  \bibinfo{author}{\bibfnamefont{G.}~\bibnamefont{Kalon}},
  \bibinfo{author}{\bibfnamefont{C.}~\bibnamefont{Bacaksiz}},
  \bibinfo{author}{\bibfnamefont{G.}~\bibnamefont{Lopez-Polin}},
  \bibinfo{author}{\bibfnamefont{T.~Y.} \bibnamefont{Zhou}},
  \bibinfo{author}{\bibfnamefont{V.}~\bibnamefont{Guarochico}},
  \bibinfo{author}{\bibfnamefont{J.}~\bibnamefont{Cai}},
  \bibinfo{author}{\bibfnamefont{C.}~\bibnamefont{Neumann}},
  \bibnamefont{et~al.}, \bibinfo{journal}{ACS Nano}
  \textbf{\bibinfo{volume}{14}}, \bibinfo{pages}{7280} (\bibinfo{year}{2020}).

\bibitem[{\citenamefont{Eldeeb et~al.}(2018)\citenamefont{Eldeeb, Fadlallah,
  Martyna, and Maarouf}}]{el18}
\bibinfo{author}{\bibfnamefont{M.~S.} \bibnamefont{Eldeeb}},
  \bibinfo{author}{\bibfnamefont{M.~M.} \bibnamefont{Fadlallah}},
  \bibinfo{author}{\bibfnamefont{G.~J.} \bibnamefont{Martyna}},
  \bibnamefont{and} \bibinfo{author}{\bibfnamefont{A.~A.}
  \bibnamefont{Maarouf}}, \bibinfo{journal}{Carbon}
  \textbf{\bibinfo{volume}{133}}, \bibinfo{pages}{369} (\bibinfo{year}{2018}).

\bibitem[{\citenamefont{Schlichting and Poulikakos}(2020)}]{sc20}
\bibinfo{author}{\bibfnamefont{K.-P.} \bibnamefont{Schlichting}}
  \bibnamefont{and}
  \bibinfo{author}{\bibfnamefont{D.}~\bibnamefont{Poulikakos}},
  \bibinfo{journal}{ACS Appl. Mater. Imterfaces} \textbf{\bibinfo{volume}{12}},
  \bibinfo{pages}{36468} (\bibinfo{year}{2020}).

\bibitem[{\citenamefont{Liu et~al.}(2021)\citenamefont{Liu, Jin, Allen, Gao,
  Ci, Kang, and Wu}}]{li21}
\bibinfo{author}{\bibfnamefont{J.}~\bibnamefont{Liu}},
  \bibinfo{author}{\bibfnamefont{L.}~\bibnamefont{Jin}},
  \bibinfo{author}{\bibfnamefont{F.}~\bibnamefont{Allen}, \bibfnamefont{I}},
  \bibinfo{author}{\bibfnamefont{Y.}~\bibnamefont{Gao}},
  \bibinfo{author}{\bibfnamefont{P.}~\bibnamefont{Ci}},
  \bibinfo{author}{\bibfnamefont{F.}~\bibnamefont{Kang}}, \bibnamefont{and}
  \bibinfo{author}{\bibfnamefont{J.}~\bibnamefont{Wu}}, \bibinfo{journal}{Nano
  Lett.} \textbf{\bibinfo{volume}{21}}, \bibinfo{pages}{2183}
  (\bibinfo{year}{2021}).

\bibitem[{\citenamefont{Liu et~al.}(2015)\citenamefont{Liu, Chen, Dai, and
  Jiang}}]{li15}
\bibinfo{author}{\bibfnamefont{H.}~\bibnamefont{Liu}},
  \bibinfo{author}{\bibfnamefont{Z.}~\bibnamefont{Chen}},
  \bibinfo{author}{\bibfnamefont{S.}~\bibnamefont{Dai}}, \bibnamefont{and}
  \bibinfo{author}{\bibfnamefont{D.}~\bibnamefont{Jiang}}, \bibinfo{journal}{J.
  Solid State Chem.} \textbf{\bibinfo{volume}{224}}, \bibinfo{pages}{2}
  (\bibinfo{year}{2015}).

\bibitem[{\citenamefont{Sun et~al.}(2019)\citenamefont{Sun, Zheng, and
  Bai}}]{su19c}
\bibinfo{author}{\bibfnamefont{C.}~\bibnamefont{Sun}},
  \bibinfo{author}{\bibfnamefont{X.}~\bibnamefont{Zheng}}, \bibnamefont{and}
  \bibinfo{author}{\bibfnamefont{B.}~\bibnamefont{Bai}},
  \bibinfo{journal}{Chem, Engin. Sci.} \textbf{\bibinfo{volume}{208}},
  \bibinfo{pages}{115141} (\bibinfo{year}{2019}).

\bibitem[{\citenamefont{Tian et~al.}(2021)\citenamefont{Tian, Duan, Luo, Cheng,
  and Shi}}]{ti21}
\bibinfo{author}{\bibfnamefont{L.}~\bibnamefont{Tian}},
  \bibinfo{author}{\bibfnamefont{H.}~\bibnamefont{Duan}},
  \bibinfo{author}{\bibfnamefont{J.}~\bibnamefont{Luo}},
  \bibinfo{author}{\bibfnamefont{Y.}~\bibnamefont{Cheng}}, \bibnamefont{and}
  \bibinfo{author}{\bibfnamefont{L.}~\bibnamefont{Shi}}, \bibinfo{journal}{ACS
  Appl. Nano Mater.} \textbf{\bibinfo{volume}{4}}, \bibinfo{pages}{9440}
  (\bibinfo{year}{2021}).

\bibitem[{\citenamefont{Lozada-Hidalgo
  et~al.}(2016)\citenamefont{Lozada-Hidalgo, Hu, Marshall, Mishchenko,
  Grigorenko, Dryfe, Radha, Grigorieva, and Geim}}]{lo16b}
\bibinfo{author}{\bibfnamefont{M.}~\bibnamefont{Lozada-Hidalgo}},
  \bibinfo{author}{\bibfnamefont{S.}~\bibnamefont{Hu}},
  \bibinfo{author}{\bibfnamefont{O.}~\bibnamefont{Marshall}},
  \bibinfo{author}{\bibfnamefont{A.}~\bibnamefont{Mishchenko}},
  \bibinfo{author}{\bibfnamefont{A.~N.} \bibnamefont{Grigorenko}},
  \bibinfo{author}{\bibfnamefont{R.~A.~W.} \bibnamefont{Dryfe}},
  \bibinfo{author}{\bibfnamefont{B.}~\bibnamefont{Radha}},
  \bibinfo{author}{\bibfnamefont{I.~V.} \bibnamefont{Grigorieva}},
  \bibnamefont{and} \bibinfo{author}{\bibfnamefont{A.~K.} \bibnamefont{Geim}},
  \bibinfo{journal}{Science} \textbf{\bibinfo{volume}{351}},
  \bibinfo{pages}{68} (\bibinfo{year}{2016}).

\bibitem[{\citenamefont{Lozada-Hidalgo
  et~al.}(2017)\citenamefont{Lozada-Hidalgo, Zhang, Hu, Esfandiar, Grigorieva,
  and Geim}}]{lo17c}
\bibinfo{author}{\bibfnamefont{M.}~\bibnamefont{Lozada-Hidalgo}},
  \bibinfo{author}{\bibfnamefont{S.}~\bibnamefont{Zhang}},
  \bibinfo{author}{\bibfnamefont{S.}~\bibnamefont{Hu}},
  \bibinfo{author}{\bibfnamefont{A.}~\bibnamefont{Esfandiar}},
  \bibinfo{author}{\bibfnamefont{I.~V.} \bibnamefont{Grigorieva}},
  \bibnamefont{and} \bibinfo{author}{\bibfnamefont{A.~K.} \bibnamefont{Geim}},
  \bibinfo{journal}{Nature Commun.} \textbf{\bibinfo{volume}{8}},
  \bibinfo{pages}{15215} (\bibinfo{year}{2017}).

\bibitem[{\citenamefont{Lu et~al.}(2009)\citenamefont{Lu, Wu, Shen, Yang, Sha,
  Cai, He, and Feng}}]{lu09b}
\bibinfo{author}{\bibfnamefont{Y.~H.} \bibnamefont{Lu}},
  \bibinfo{author}{\bibfnamefont{R.~Q.} \bibnamefont{Wu}},
  \bibinfo{author}{\bibfnamefont{L.}~\bibnamefont{Shen}},
  \bibinfo{author}{\bibfnamefont{M.}~\bibnamefont{Yang}},
  \bibinfo{author}{\bibfnamefont{Z.~D.} \bibnamefont{Sha}},
  \bibinfo{author}{\bibfnamefont{Y.~Q.} \bibnamefont{Cai}},
  \bibinfo{author}{\bibfnamefont{P.~M.} \bibnamefont{He}}, \bibnamefont{and}
  \bibinfo{author}{\bibfnamefont{Y.~P.} \bibnamefont{Feng}},
  \bibinfo{journal}{Appl. Phys. Lett} \textbf{\bibinfo{volume}{94}},
  \bibinfo{pages}{122111} (\bibinfo{year}{2009}).

\bibitem[{\citenamefont{Islam et~al.}(2014)\citenamefont{Islam, Tanaka, and
  Hashimoto}}]{is14}
\bibinfo{author}{\bibfnamefont{M.~S.} \bibnamefont{Islam}},
  \bibinfo{author}{\bibfnamefont{S.}~\bibnamefont{Tanaka}}, \bibnamefont{and}
  \bibinfo{author}{\bibfnamefont{A.}~\bibnamefont{Hashimoto}},
  \bibinfo{journal}{Carbon} \textbf{\bibinfo{volume}{80}}, \bibinfo{pages}{146}
  (\bibinfo{year}{2014}).

\bibitem[{\citenamefont{Patera et~al.}(2015)\citenamefont{Patera, Bianchini,
  Troiano, Dri, Cepek, Peressi, Africh, and Comelli}}]{pa14b}
\bibinfo{author}{\bibfnamefont{L.~L.} \bibnamefont{Patera}},
  \bibinfo{author}{\bibfnamefont{F.}~\bibnamefont{Bianchini}},
  \bibinfo{author}{\bibfnamefont{G.}~\bibnamefont{Troiano}},
  \bibinfo{author}{\bibfnamefont{C.}~\bibnamefont{Dri}},
  \bibinfo{author}{\bibfnamefont{C.}~\bibnamefont{Cepek}},
  \bibinfo{author}{\bibfnamefont{M.}~\bibnamefont{Peressi}},
  \bibinfo{author}{\bibfnamefont{C.}~\bibnamefont{Africh}}, \bibnamefont{and}
  \bibinfo{author}{\bibfnamefont{G.}~\bibnamefont{Comelli}},
  \bibinfo{journal}{Nano Lett.} \textbf{\bibinfo{volume}{15}},
  \bibinfo{pages}{56} (\bibinfo{year}{2015}).

\bibitem[{\citenamefont{Sluiter and Kawazoe}(2003)}]{sl03}
\bibinfo{author}{\bibfnamefont{M.~H.~F.} \bibnamefont{Sluiter}}
  \bibnamefont{and} \bibinfo{author}{\bibfnamefont{Y.}~\bibnamefont{Kawazoe}},
  \bibinfo{journal}{Phys. Rev. B} \textbf{\bibinfo{volume}{68}},
  \bibinfo{pages}{085410} (\bibinfo{year}{2003}).

\bibitem[{\citenamefont{Hornek{\ae}r et~al.}(2006)\citenamefont{Hornek{\ae}r,
  Rauls, Xu, Sljivancanin, Otero, Stensgaard, Laegsgaard, Hammer, and
  Besenbacher}}]{ho06b}
\bibinfo{author}{\bibfnamefont{L.}~\bibnamefont{Hornek{\ae}r}},
  \bibinfo{author}{\bibfnamefont{E.}~\bibnamefont{Rauls}},
  \bibinfo{author}{\bibfnamefont{W.}~\bibnamefont{Xu}},
  \bibinfo{author}{\bibfnamefont{Z.}~\bibnamefont{Sljivancanin}},
  \bibinfo{author}{\bibfnamefont{R.}~\bibnamefont{Otero}},
  \bibinfo{author}{\bibfnamefont{I.}~\bibnamefont{Stensgaard}},
  \bibinfo{author}{\bibfnamefont{E.}~\bibnamefont{Laegsgaard}},
  \bibinfo{author}{\bibfnamefont{B.}~\bibnamefont{Hammer}}, \bibnamefont{and}
  \bibinfo{author}{\bibfnamefont{F.}~\bibnamefont{Besenbacher}},
  \bibinfo{journal}{Phys. Rev. Lett.} \textbf{\bibinfo{volume}{97}},
  \bibinfo{pages}{186102} (\bibinfo{year}{2006}).

\bibitem[{\citenamefont{Casolo et~al.}(2009)\citenamefont{Casolo, Lovvik,
  Martinazzo, and Tantardini}}]{ca09}
\bibinfo{author}{\bibfnamefont{S.}~\bibnamefont{Casolo}},
  \bibinfo{author}{\bibfnamefont{O.~M.} \bibnamefont{Lovvik}},
  \bibinfo{author}{\bibfnamefont{R.}~\bibnamefont{Martinazzo}},
  \bibnamefont{and} \bibinfo{author}{\bibfnamefont{G.~F.}
  \bibnamefont{Tantardini}}, \bibinfo{journal}{J. Chem. Phys.}
  \textbf{\bibinfo{volume}{130}}, \bibinfo{pages}{054704}
  (\bibinfo{year}{2009}).

\bibitem[{\citenamefont{Boukhvalov et~al.}(2008)\citenamefont{Boukhvalov,
  Katsnelson, and Lichtenstein}}]{bo08}
\bibinfo{author}{\bibfnamefont{D.~W.} \bibnamefont{Boukhvalov}},
  \bibinfo{author}{\bibfnamefont{M.~I.} \bibnamefont{Katsnelson}},
  \bibnamefont{and} \bibinfo{author}{\bibfnamefont{A.~I.}
  \bibnamefont{Lichtenstein}}, \bibinfo{journal}{Phys. Rev. B}
  \textbf{\bibinfo{volume}{77}}, \bibinfo{pages}{035427}
  (\bibinfo{year}{2008}).

\bibitem[{\citenamefont{Duplock et~al.}(2004)\citenamefont{Duplock, Scheffler,
  and Lindan}}]{du04}
\bibinfo{author}{\bibfnamefont{E.~J.} \bibnamefont{Duplock}},
  \bibinfo{author}{\bibfnamefont{M.}~\bibnamefont{Scheffler}},
  \bibnamefont{and} \bibinfo{author}{\bibfnamefont{P.~J.~D.}
  \bibnamefont{Lindan}}, \bibinfo{journal}{Phys. Rev. Lett.}
  \textbf{\bibinfo{volume}{92}}, \bibinfo{pages}{225502}
  (\bibinfo{year}{2004}).

\bibitem[{\citenamefont{de~Andres and Verg\'es}(2008)}]{an08}
\bibinfo{author}{\bibfnamefont{P.~L.} \bibnamefont{de~Andres}}
  \bibnamefont{and} \bibinfo{author}{\bibfnamefont{J.~A.}
  \bibnamefont{Verg\'es}}, \bibinfo{journal}{Appl. Phys. Lett.}
  \textbf{\bibinfo{volume}{93}}, \bibinfo{pages}{171915}
  (\bibinfo{year}{2008}).

\bibitem[{\citenamefont{Despiau-Pujo et~al.}(2013)\citenamefont{Despiau-Pujo,
  Davydova, Cunge, Delfour, Magaud, and Graves}}]{de13}
\bibinfo{author}{\bibfnamefont{E.}~\bibnamefont{Despiau-Pujo}},
  \bibinfo{author}{\bibfnamefont{A.}~\bibnamefont{Davydova}},
  \bibinfo{author}{\bibfnamefont{G.}~\bibnamefont{Cunge}},
  \bibinfo{author}{\bibfnamefont{L.}~\bibnamefont{Delfour}},
  \bibinfo{author}{\bibfnamefont{L.}~\bibnamefont{Magaud}}, \bibnamefont{and}
  \bibinfo{author}{\bibfnamefont{D.~B.} \bibnamefont{Graves}},
  \bibinfo{journal}{J. Appl. Phys.} \textbf{\bibinfo{volume}{113}},
  \bibinfo{pages}{114302} (\bibinfo{year}{2013}).

\bibitem[{\citenamefont{Bonfanti et~al.}(2018)\citenamefont{Bonfanti, Achilli,
  and Martinazzo}}]{bo18b}
\bibinfo{author}{\bibfnamefont{M.}~\bibnamefont{Bonfanti}},
  \bibinfo{author}{\bibfnamefont{S.}~\bibnamefont{Achilli}}, \bibnamefont{and}
  \bibinfo{author}{\bibfnamefont{R.}~\bibnamefont{Martinazzo}},
  \bibinfo{journal}{J. Phys.: Condens. Matter} \textbf{\bibinfo{volume}{30}},
  \bibinfo{pages}{283002} (\bibinfo{year}{2018}).

\bibitem[{\citenamefont{Petucci et~al.}(2013)\citenamefont{Petucci, LeBlond,
  Karimi, and Vidali}}]{pe13b}
\bibinfo{author}{\bibfnamefont{J.}~\bibnamefont{Petucci}},
  \bibinfo{author}{\bibfnamefont{C.}~\bibnamefont{LeBlond}},
  \bibinfo{author}{\bibfnamefont{M.}~\bibnamefont{Karimi}}, \bibnamefont{and}
  \bibinfo{author}{\bibfnamefont{G.}~\bibnamefont{Vidali}},
  \bibinfo{journal}{J. Chem. Phys.} \textbf{\bibinfo{volume}{139}},
  \bibinfo{pages}{044706} (\bibinfo{year}{2013}).

\bibitem[{\citenamefont{Petucci et~al.}(2018)\citenamefont{Petucci, Semone,
  LeBlond, Karimi, and Vidali}}]{pe18b}
\bibinfo{author}{\bibfnamefont{J.}~\bibnamefont{Petucci}},
  \bibinfo{author}{\bibfnamefont{S.}~\bibnamefont{Semone}},
  \bibinfo{author}{\bibfnamefont{C.}~\bibnamefont{LeBlond}},
  \bibinfo{author}{\bibfnamefont{M.}~\bibnamefont{Karimi}}, \bibnamefont{and}
  \bibinfo{author}{\bibfnamefont{G.}~\bibnamefont{Vidali}},
  \bibinfo{journal}{J. Chem. Phys.} \textbf{\bibinfo{volume}{149}},
  \bibinfo{pages}{014702} (\bibinfo{year}{2018}).

\bibitem[{\citenamefont{Balog et~al.}(2009)\citenamefont{Balog, Jorgensen,
  Wells, Laegsgaard, Hofmann, Besenbacher, and Hornek{\ae}r}}]{ba09b}
\bibinfo{author}{\bibfnamefont{R.}~\bibnamefont{Balog}},
  \bibinfo{author}{\bibfnamefont{B.}~\bibnamefont{Jorgensen}},
  \bibinfo{author}{\bibfnamefont{J.}~\bibnamefont{Wells}},
  \bibinfo{author}{\bibfnamefont{E.}~\bibnamefont{Laegsgaard}},
  \bibinfo{author}{\bibfnamefont{P.}~\bibnamefont{Hofmann}},
  \bibinfo{author}{\bibfnamefont{F.}~\bibnamefont{Besenbacher}},
  \bibnamefont{and}
  \bibinfo{author}{\bibfnamefont{L.}~\bibnamefont{Hornek{\ae}r}},
  \bibinfo{journal}{J. Am. Chem. Soc.} \textbf{\bibinfo{volume}{131}},
  \bibinfo{pages}{8744} (\bibinfo{year}{2009}).

\bibitem[{\citenamefont{Balog et~al.}(2010)\citenamefont{Balog, Jorgensen,
  Nilsson, Andersen, Rienks, Bianchi, Fanetti, Laegsgaard, Baraldi, Lizzit
  et~al.}}]{ba10}
\bibinfo{author}{\bibfnamefont{R.}~\bibnamefont{Balog}},
  \bibinfo{author}{\bibfnamefont{B.}~\bibnamefont{Jorgensen}},
  \bibinfo{author}{\bibfnamefont{L.}~\bibnamefont{Nilsson}},
  \bibinfo{author}{\bibfnamefont{M.}~\bibnamefont{Andersen}},
  \bibinfo{author}{\bibfnamefont{E.}~\bibnamefont{Rienks}},
  \bibinfo{author}{\bibfnamefont{M.}~\bibnamefont{Bianchi}},
  \bibinfo{author}{\bibfnamefont{M.}~\bibnamefont{Fanetti}},
  \bibinfo{author}{\bibfnamefont{E.}~\bibnamefont{Laegsgaard}},
  \bibinfo{author}{\bibfnamefont{A.}~\bibnamefont{Baraldi}},
  \bibinfo{author}{\bibfnamefont{S.}~\bibnamefont{Lizzit}},
  \bibnamefont{et~al.}, \bibinfo{journal}{Nature Mater.}
  \textbf{\bibinfo{volume}{9}}, \bibinfo{pages}{315} (\bibinfo{year}{2010}).

\bibitem[{\citenamefont{Ulstrup et~al.}(2013)\citenamefont{Ulstrup, Nilsson,
  Miwa, Balog, Bianchi, Hornek{\ae}r, and Hofmann}}]{ul13}
\bibinfo{author}{\bibfnamefont{S.}~\bibnamefont{Ulstrup}},
  \bibinfo{author}{\bibfnamefont{L.}~\bibnamefont{Nilsson}},
  \bibinfo{author}{\bibfnamefont{J.~A.} \bibnamefont{Miwa}},
  \bibinfo{author}{\bibfnamefont{R.}~\bibnamefont{Balog}},
  \bibinfo{author}{\bibfnamefont{M.}~\bibnamefont{Bianchi}},
  \bibinfo{author}{\bibfnamefont{L.}~\bibnamefont{Hornek{\ae}r}},
  \bibnamefont{and} \bibinfo{author}{\bibfnamefont{P.}~\bibnamefont{Hofmann}},
  \bibinfo{journal}{Phys. Rev. B} \textbf{\bibinfo{volume}{88}},
  \bibinfo{pages}{125425} (\bibinfo{year}{2013}).

\bibitem[{\citenamefont{Herrero and Ram\'irez}(2010)}]{he10b}
\bibinfo{author}{\bibfnamefont{C.~P.} \bibnamefont{Herrero}} \bibnamefont{and}
  \bibinfo{author}{\bibfnamefont{R.}~\bibnamefont{Ram\'irez}},
  \bibinfo{journal}{J. Phys. D: Appl. Phys.} \textbf{\bibinfo{volume}{43}},
  \bibinfo{pages}{255402} (\bibinfo{year}{2010}).

\bibitem[{\citenamefont{Herrero and Ram\'irez}(2009)}]{he09a}
\bibinfo{author}{\bibfnamefont{C.~P.} \bibnamefont{Herrero}} \bibnamefont{and}
  \bibinfo{author}{\bibfnamefont{R.}~\bibnamefont{Ram\'irez}},
  \bibinfo{journal}{Phys. Rev. B} \textbf{\bibinfo{volume}{79}},
  \bibinfo{pages}{115429} (\bibinfo{year}{2009}).

\bibitem[{\citenamefont{Herrero and Ram\'{\i}rez}(2007)}]{he07}
\bibinfo{author}{\bibfnamefont{C.~P.} \bibnamefont{Herrero}} \bibnamefont{and}
  \bibinfo{author}{\bibfnamefont{R.}~\bibnamefont{Ram\'{\i}rez}},
  \bibinfo{journal}{Phys. Rev. Lett.} \textbf{\bibinfo{volume}{99}},
  \bibinfo{pages}{205504} (\bibinfo{year}{2007}).

\bibitem[{\citenamefont{Panzarini and Colombo}(1994)}]{pa94}
\bibinfo{author}{\bibfnamefont{G.}~\bibnamefont{Panzarini}} \bibnamefont{and}
  \bibinfo{author}{\bibfnamefont{L.}~\bibnamefont{Colombo}},
  \bibinfo{journal}{Phys. Rev. Lett.} \textbf{\bibinfo{volume}{73}},
  \bibinfo{pages}{1636} (\bibinfo{year}{1994}).

\bibitem[{\citenamefont{B\'edard and Lewis}(2000)}]{be00}
\bibinfo{author}{\bibfnamefont{S.}~\bibnamefont{B\'edard}} \bibnamefont{and}
  \bibinfo{author}{\bibfnamefont{L.~J.} \bibnamefont{Lewis}},
  \bibinfo{journal}{Phys. Rev. B} \textbf{\bibinfo{volume}{61}},
  \bibinfo{pages}{9895} (\bibinfo{year}{2000}).

\bibitem[{\citenamefont{Boucher and DeLeo}(1994)}]{bo94}
\bibinfo{author}{\bibfnamefont{D.~E.} \bibnamefont{Boucher}} \bibnamefont{and}
  \bibinfo{author}{\bibfnamefont{G.~G.} \bibnamefont{DeLeo}},
  \bibinfo{journal}{Phys. Rev. B} \textbf{\bibinfo{volume}{50}},
  \bibinfo{pages}{5247} (\bibinfo{year}{1994}).

\bibitem[{\citenamefont{Shabaev et~al.}(2010)\citenamefont{Shabaev,
  Papaconstantopoulos, Mehl, and Bernstein}}]{sh10}
\bibinfo{author}{\bibfnamefont{A.}~\bibnamefont{Shabaev}},
  \bibinfo{author}{\bibfnamefont{D.~A.} \bibnamefont{Papaconstantopoulos}},
  \bibinfo{author}{\bibfnamefont{M.~J.} \bibnamefont{Mehl}}, \bibnamefont{and}
  \bibinfo{author}{\bibfnamefont{N.}~\bibnamefont{Bernstein}},
  \bibinfo{journal}{Phys. Rev. B} \textbf{\bibinfo{volume}{81}},
  \bibinfo{pages}{184103} (\bibinfo{year}{2010}).

\bibitem[{\citenamefont{Ukpong}(2010)}]{uk10}
\bibinfo{author}{\bibfnamefont{A.~M.} \bibnamefont{Ukpong}},
  \bibinfo{journal}{Mol. Phys.} \textbf{\bibinfo{volume}{108}},
  \bibinfo{pages}{1607} (\bibinfo{year}{2010}).

\bibitem[{\citenamefont{Hayashi et~al.}(2011)\citenamefont{Hayashi, Tezuka,
  Ozawa, Shimazaki, Adachi, and Kubo}}]{ha11}
\bibinfo{author}{\bibfnamefont{K.}~\bibnamefont{Hayashi}},
  \bibinfo{author}{\bibfnamefont{K.}~\bibnamefont{Tezuka}},
  \bibinfo{author}{\bibfnamefont{N.}~\bibnamefont{Ozawa}},
  \bibinfo{author}{\bibfnamefont{T.}~\bibnamefont{Shimazaki}},
  \bibinfo{author}{\bibfnamefont{K.}~\bibnamefont{Adachi}}, \bibnamefont{and}
  \bibinfo{author}{\bibfnamefont{M.}~\bibnamefont{Kubo}}, \bibinfo{journal}{J.
  Phys. Chem. C} \textbf{\bibinfo{volume}{115}}, \bibinfo{pages}{22981}
  (\bibinfo{year}{2011}).

\bibitem[{\citenamefont{Dominguez-Gutierrez
  et~al.}(2018)\citenamefont{Dominguez-Gutierrez, Krstic, Irle, and
  Cabrera-Trujillo}}]{do18}
\bibinfo{author}{\bibfnamefont{F.~J.} \bibnamefont{Dominguez-Gutierrez}},
  \bibinfo{author}{\bibfnamefont{P.~S.} \bibnamefont{Krstic}},
  \bibinfo{author}{\bibfnamefont{S.}~\bibnamefont{Irle}}, \bibnamefont{and}
  \bibinfo{author}{\bibfnamefont{R.}~\bibnamefont{Cabrera-Trujillo}},
  \bibinfo{journal}{Carbon} \textbf{\bibinfo{volume}{134}},
  \bibinfo{pages}{189} (\bibinfo{year}{2018}).

\bibitem[{\citenamefont{Mananghaya et~al.}(2018)\citenamefont{Mananghaya,
  Santos, and Yu}}]{ma18c}
\bibinfo{author}{\bibfnamefont{M.~R.} \bibnamefont{Mananghaya}},
  \bibinfo{author}{\bibfnamefont{G.~N.} \bibnamefont{Santos}},
  \bibnamefont{and} \bibinfo{author}{\bibfnamefont{D.}~\bibnamefont{Yu}},
  \bibinfo{journal}{Adsorption} \textbf{\bibinfo{volume}{24}},
  \bibinfo{pages}{683} (\bibinfo{year}{2018}).

\bibitem[{\citenamefont{Dominguez-Gutierrez
  et~al.}(2019)\citenamefont{Dominguez-Gutierrez, Martinez-Flores, and
  Cabrera-Trujillo}}]{do19}
\bibinfo{author}{\bibfnamefont{F.~J.} \bibnamefont{Dominguez-Gutierrez}},
  \bibinfo{author}{\bibfnamefont{C.}~\bibnamefont{Martinez-Flores}},
  \bibnamefont{and}
  \bibinfo{author}{\bibfnamefont{R.}~\bibnamefont{Cabrera-Trujillo}},
  \bibinfo{journal}{J. Appl. Phys.} \textbf{\bibinfo{volume}{125}},
  \bibinfo{pages}{094506} (\bibinfo{year}{2019}).

\bibitem[{\citenamefont{Porezag et~al.}(1995)\citenamefont{Porezag, Frauenheim,
  K\"ohler, Seifert, and Kaschner}}]{po95}
\bibinfo{author}{\bibfnamefont{D.}~\bibnamefont{Porezag}},
  \bibinfo{author}{\bibfnamefont{T.}~\bibnamefont{Frauenheim}},
  \bibinfo{author}{\bibfnamefont{T.}~\bibnamefont{K\"ohler}},
  \bibinfo{author}{\bibfnamefont{G.}~\bibnamefont{Seifert}}, \bibnamefont{and}
  \bibinfo{author}{\bibfnamefont{R.}~\bibnamefont{Kaschner}},
  \bibinfo{journal}{Phys. Rev. B} \textbf{\bibinfo{volume}{51}},
  \bibinfo{pages}{12947} (\bibinfo{year}{1995}).

\bibitem[{\citenamefont{Goringe et~al.}(1997)\citenamefont{Goringe, Bowler, and
  Hern\'andez}}]{go97}
\bibinfo{author}{\bibfnamefont{C.~M.} \bibnamefont{Goringe}},
  \bibinfo{author}{\bibfnamefont{D.~R.} \bibnamefont{Bowler}},
  \bibnamefont{and}
  \bibinfo{author}{\bibfnamefont{E.}~\bibnamefont{Hern\'andez}},
  \bibinfo{journal}{Rep. Prog. Phys.} \textbf{\bibinfo{volume}{60}},
  \bibinfo{pages}{1447} (\bibinfo{year}{1997}).

\bibitem[{\citenamefont{Colombo}(2005)}]{co05}
\bibinfo{author}{\bibfnamefont{L.}~\bibnamefont{Colombo}},
  \bibinfo{journal}{Riv. Nuovo Cimento} \textbf{\bibinfo{volume}{28}},
  \bibinfo{pages}{1} (\bibinfo{year}{2005}).

\bibitem[{\citenamefont{Herrero et~al.}(2006)\citenamefont{Herrero,
  Ram\'{\i}rez, and Hern\'andez}}]{he06}
\bibinfo{author}{\bibfnamefont{C.~P.} \bibnamefont{Herrero}},
  \bibinfo{author}{\bibfnamefont{R.}~\bibnamefont{Ram\'{\i}rez}},
  \bibnamefont{and} \bibinfo{author}{\bibfnamefont{E.~R.}
  \bibnamefont{Hern\'andez}}, \bibinfo{journal}{Phys. Rev. B}
  \textbf{\bibinfo{volume}{73}}, \bibinfo{pages}{245211}
  (\bibinfo{year}{2006}).

\bibitem[{\citenamefont{Johnson et~al.}(1993)\citenamefont{Johnson, Gill, and
  Pople}}]{jo93}
\bibinfo{author}{\bibfnamefont{B.~G.} \bibnamefont{Johnson}},
  \bibinfo{author}{\bibfnamefont{P.~M.~W.} \bibnamefont{Gill}},
  \bibnamefont{and} \bibinfo{author}{\bibfnamefont{J.~A.} \bibnamefont{Pople}},
  \bibinfo{journal}{J. Chem. Phys.} \textbf{\bibinfo{volume}{98}},
  \bibinfo{pages}{5612} (\bibinfo{year}{1993}).

\bibitem[{\citenamefont{Herrero and Ram\'{\i}rez}(2010)}]{he10}
\bibinfo{author}{\bibfnamefont{C.~P.} \bibnamefont{Herrero}} \bibnamefont{and}
  \bibinfo{author}{\bibfnamefont{R.}~\bibnamefont{Ram\'{\i}rez}},
  \bibinfo{journal}{Phys. Rev. B} \textbf{\bibinfo{volume}{82}},
  \bibinfo{pages}{174117} (\bibinfo{year}{2010}).

\bibitem[{\citenamefont{Tuckerman and Hughes}(1998)}]{tu98}
\bibinfo{author}{\bibfnamefont{M.~E.} \bibnamefont{Tuckerman}}
  \bibnamefont{and} \bibinfo{author}{\bibfnamefont{A.}~\bibnamefont{Hughes}},
  in \emph{\bibinfo{booktitle}{Classical and Quantum Dynamics in Condensed
  Phase Simulations}}, edited by \bibinfo{editor}{\bibfnamefont{B.~J.}
  \bibnamefont{Berne}},
  \bibinfo{editor}{\bibfnamefont{G.}~\bibnamefont{Ciccotti}}, \bibnamefont{and}
  \bibinfo{editor}{\bibfnamefont{D.~F.} \bibnamefont{Coker}}
  (\bibinfo{publisher}{Word Scientific}, \bibinfo{address}{Singapore},
  \bibinfo{year}{1998}), p. \bibinfo{pages}{311}.

\bibitem[{\citenamefont{Allen and Tildesley}(1987)}]{al87}
\bibinfo{author}{\bibfnamefont{M.~P.} \bibnamefont{Allen}} \bibnamefont{and}
  \bibinfo{author}{\bibfnamefont{D.~J.} \bibnamefont{Tildesley}},
  \emph{\bibinfo{title}{Computer simulation of liquids}}
  (\bibinfo{publisher}{Clarendon Press}, \bibinfo{address}{Oxford},
  \bibinfo{year}{1987}).

\bibitem[{\citenamefont{Martyna et~al.}(1996)\citenamefont{Martyna, Tuckerman,
  Tobias, and Klein}}]{ma96}
\bibinfo{author}{\bibfnamefont{G.~J.} \bibnamefont{Martyna}},
  \bibinfo{author}{\bibfnamefont{M.~E.} \bibnamefont{Tuckerman}},
  \bibinfo{author}{\bibfnamefont{D.~J.} \bibnamefont{Tobias}},
  \bibnamefont{and} \bibinfo{author}{\bibfnamefont{M.~L.} \bibnamefont{Klein}},
  \bibinfo{journal}{Mol. Phys.} \textbf{\bibinfo{volume}{87}},
  \bibinfo{pages}{1117} (\bibinfo{year}{1996}).

\bibitem[{\citenamefont{{Frisch \it et al.}}(2016)}]{ga16}
\bibinfo{author}{\bibfnamefont{M.~J.} \bibnamefont{{Frisch \it et al.}}},
  \emph{\bibinfo{title}{Gaussian 16, Revision A.03}}
  (\bibinfo{publisher}{Gaussian Inc.}, \bibinfo{address}{Wallingford, CT},
  \bibinfo{year}{2016}).

\bibitem[{\citenamefont{Becke}(1993)}]{be93}
\bibinfo{author}{\bibfnamefont{A.}~\bibnamefont{Becke}}, \bibinfo{journal}{J.
  Chem. Phys.} \textbf{\bibinfo{volume}{98}}, \bibinfo{pages}{5648}
  (\bibinfo{year}{1993}).

\bibitem[{\citenamefont{Weigend and Ahlrichs}(2005)}]{we05}
\bibinfo{author}{\bibfnamefont{F.}~\bibnamefont{Weigend}} \bibnamefont{and}
  \bibinfo{author}{\bibfnamefont{R.}~\bibnamefont{Ahlrichs}},
  \bibinfo{journal}{Phys. Chem. Chem. Phys.} \textbf{\bibinfo{volume}{7}},
  \bibinfo{pages}{3297} (\bibinfo{year}{2005}).

\bibitem[{\citenamefont{Li and Frisch}(2006)}]{li06}
\bibinfo{author}{\bibfnamefont{X.}~\bibnamefont{Li}} \bibnamefont{and}
  \bibinfo{author}{\bibfnamefont{M.~J.} \bibnamefont{Frisch}},
  \bibinfo{journal}{J. Chem. Theory Comput.} \textbf{\bibinfo{volume}{2}},
  \bibinfo{pages}{835} (\bibinfo{year}{2006}).

\bibitem[{\citenamefont{Verg\'es and de~Andres}(2010)}]{ve10}
\bibinfo{author}{\bibfnamefont{J.~A.} \bibnamefont{Verg\'es}} \bibnamefont{and}
  \bibinfo{author}{\bibfnamefont{P.~L.} \bibnamefont{de~Andres}},
  \bibinfo{journal}{Phys. Rev. B} \textbf{\bibinfo{volume}{81}},
  \bibinfo{pages}{075423} (\bibinfo{year}{2010}).

\bibitem[{\citenamefont{Yazyev and Helm}(2007)}]{ya07}
\bibinfo{author}{\bibfnamefont{O.~V.} \bibnamefont{Yazyev}} \bibnamefont{and}
  \bibinfo{author}{\bibfnamefont{L.}~\bibnamefont{Helm}},
  \bibinfo{journal}{Phys. Rev. B} \textbf{\bibinfo{volume}{75}},
  \bibinfo{pages}{125408} (\bibinfo{year}{2007}).

\bibitem[{\citenamefont{McKay et~al.}(2010)\citenamefont{McKay, Wales, Jenkins,
  Verg\'es, and de~Andres}}]{mc10}
\bibinfo{author}{\bibfnamefont{H.}~\bibnamefont{McKay}},
  \bibinfo{author}{\bibfnamefont{D.~J.} \bibnamefont{Wales}},
  \bibinfo{author}{\bibfnamefont{S.~J.} \bibnamefont{Jenkins}},
  \bibinfo{author}{\bibfnamefont{J.~A.} \bibnamefont{Verg\'es}},
  \bibnamefont{and} \bibinfo{author}{\bibfnamefont{P.~L.}
  \bibnamefont{de~Andres}}, \bibinfo{journal}{Phys. Rev. B}
  \textbf{\bibinfo{volume}{81}}, \bibinfo{pages}{075425}
  (\bibinfo{year}{2010}).

\bibitem[{\citenamefont{Ram\'irez and L\'opez-Ciudad}(2001)}]{ra01}
\bibinfo{author}{\bibfnamefont{R.}~\bibnamefont{Ram\'irez}} \bibnamefont{and}
  \bibinfo{author}{\bibfnamefont{T.}~\bibnamefont{L\'opez-Ciudad}},
  \bibinfo{journal}{J. Chem. Phys.} \textbf{\bibinfo{volume}{115}},
  \bibinfo{pages}{103} (\bibinfo{year}{2001}).

\bibitem[{\citenamefont{Ram\'irez and Herrero}(2019)}]{ra19}
\bibinfo{author}{\bibfnamefont{R.}~\bibnamefont{Ram\'irez}} \bibnamefont{and}
  \bibinfo{author}{\bibfnamefont{C.~P.} \bibnamefont{Herrero}},
  \bibinfo{journal}{J. Chem. Phys.} \textbf{\bibinfo{volume}{151}},
  \bibinfo{pages}{224107} (\bibinfo{year}{2019}).

\bibitem[{\citenamefont{Wheeler et~al.}(2003)\citenamefont{Wheeler, Dong, and
  Boesch}}]{wh03}
\bibinfo{author}{\bibfnamefont{R.~A.} \bibnamefont{Wheeler}},
  \bibinfo{author}{\bibfnamefont{H.}~\bibnamefont{Dong}}, \bibnamefont{and}
  \bibinfo{author}{\bibfnamefont{S.~E.} \bibnamefont{Boesch}},
  \bibinfo{journal}{ChemPhysChem} \textbf{\bibinfo{volume}{4}},
  \bibinfo{pages}{382} (\bibinfo{year}{2003}).

\bibitem[{\citenamefont{Schmitz and Tavan}(2004)}]{sc04}
\bibinfo{author}{\bibfnamefont{M.}~\bibnamefont{Schmitz}} \bibnamefont{and}
  \bibinfo{author}{\bibfnamefont{P.}~\bibnamefont{Tavan}}, \bibinfo{journal}{J.
  Chem. Phys.} \textbf{\bibinfo{volume}{121}}, \bibinfo{pages}{12233}
  (\bibinfo{year}{2004}).

\bibitem[{\citenamefont{Ram\'irez and Herrero}(2020)}]{ra20}
\bibinfo{author}{\bibfnamefont{R.}~\bibnamefont{Ram\'irez}} \bibnamefont{and}
  \bibinfo{author}{\bibfnamefont{C.~P.} \bibnamefont{Herrero}},
  \bibinfo{journal}{Phys. Rev. B} \textbf{\bibinfo{volume}{101}},
  \bibinfo{pages}{235436} (\bibinfo{year}{2020}).

\bibitem[{\citenamefont{L\'opez-Ciudad
  et~al.}(2003)\citenamefont{L\'opez-Ciudad, Ram\'irez, Schulte, and
  B\"ohm}}]{lo03}
\bibinfo{author}{\bibfnamefont{T.}~\bibnamefont{L\'opez-Ciudad}},
  \bibinfo{author}{\bibfnamefont{R.}~\bibnamefont{Ram\'irez}},
  \bibinfo{author}{\bibfnamefont{J.}~\bibnamefont{Schulte}}, \bibnamefont{and}
  \bibinfo{author}{\bibfnamefont{M.~C.} \bibnamefont{B\"ohm}},
  \bibinfo{journal}{J. Chem. Phys.} \textbf{\bibinfo{volume}{119}},
  \bibinfo{pages}{4328} (\bibinfo{year}{2003}).

\bibitem[{\citenamefont{Casartelli et~al.}(2014)\citenamefont{Casartelli,
  Casolo, Tantardini, and Martinazzo}}]{ca14}
\bibinfo{author}{\bibfnamefont{M.}~\bibnamefont{Casartelli}},
  \bibinfo{author}{\bibfnamefont{S.}~\bibnamefont{Casolo}},
  \bibinfo{author}{\bibfnamefont{G.~F.} \bibnamefont{Tantardini}},
  \bibnamefont{and}
  \bibinfo{author}{\bibfnamefont{R.}~\bibnamefont{Martinazzo}},
  \bibinfo{journal}{Carbon} \textbf{\bibinfo{volume}{77}}, \bibinfo{pages}{165}
  (\bibinfo{year}{2014}).

\bibitem[{\citenamefont{Wirtz and Rubio}(2004)}]{wi04}
\bibinfo{author}{\bibfnamefont{L.}~\bibnamefont{Wirtz}} \bibnamefont{and}
  \bibinfo{author}{\bibfnamefont{A.}~\bibnamefont{Rubio}},
  \bibinfo{journal}{Solid State Commun.} \textbf{\bibinfo{volume}{131}},
  \bibinfo{pages}{141} (\bibinfo{year}{2004}).

\bibitem[{\citenamefont{Bartolomei et~al.}(2021)\citenamefont{Bartolomei,
  Hernandez, Campos-Martinez, Hernandez-Lamoneda, and Giorgi}}]{ba21}
\bibinfo{author}{\bibfnamefont{M.}~\bibnamefont{Bartolomei}},
  \bibinfo{author}{\bibfnamefont{M.}~\bibnamefont{Hernandez},
  \bibfnamefont{I}},
  \bibinfo{author}{\bibfnamefont{J.}~\bibnamefont{Campos-Martinez}},
  \bibinfo{author}{\bibfnamefont{R.}~\bibnamefont{Hernandez-Lamoneda}},
  \bibnamefont{and} \bibinfo{author}{\bibfnamefont{G.}~\bibnamefont{Giorgi}},
  \bibinfo{journal}{Carbon} \textbf{\bibinfo{volume}{178}},
  \bibinfo{pages}{718} (\bibinfo{year}{2021}).

\bibitem[{\citenamefont{Philibert}(1991)}]{ph91}
\bibinfo{author}{\bibfnamefont{J.}~\bibnamefont{Philibert}},
  \emph{\bibinfo{title}{Atom movements. Diffusion and transport in solids}}
  (\bibinfo{publisher}{EDP Sciences}, \bibinfo{address}{Les Ulis, France},
  \bibinfo{year}{1991}).

\bibitem[{\citenamefont{Feller}(1968)}]{fe68}
\bibinfo{author}{\bibfnamefont{W.}~\bibnamefont{Feller}},
  \emph{\bibinfo{title}{An Introduction to Probability Theory and
  Applications}}, vol.~\bibinfo{volume}{1} (\bibinfo{publisher}{Wiley},
  \bibinfo{address}{New York}, \bibinfo{year}{1968}), \bibinfo{edition}{3rd}
  ed.

\bibitem[{\citenamefont{Feldman and Valdez-Flores}(2010)}]{fe10}
\bibinfo{author}{\bibfnamefont{R.~M.} \bibnamefont{Feldman}} \bibnamefont{and}
  \bibinfo{author}{\bibfnamefont{C.}~\bibnamefont{Valdez-Flores}},
  \emph{\bibinfo{title}{Applied Probability and Stochastic Processes}}
  (\bibinfo{publisher}{Springer}, \bibinfo{address}{Heidelberg},
  \bibinfo{year}{2010}), \bibinfo{edition}{2nd} ed.

\bibitem[{\citenamefont{Flynn and Stoneham}(1970)}]{fl70}
\bibinfo{author}{\bibfnamefont{C.~P.} \bibnamefont{Flynn}} \bibnamefont{and}
  \bibinfo{author}{\bibfnamefont{A.~M.} \bibnamefont{Stoneham}},
  \bibinfo{journal}{Phys. Rev. B} \textbf{\bibinfo{volume}{1}},
  \bibinfo{pages}{3966} (\bibinfo{year}{1970}).

\bibitem[{\citenamefont{Sugimoto and Fukai}(1980)}]{su80}
\bibinfo{author}{\bibfnamefont{H.}~\bibnamefont{Sugimoto}} \bibnamefont{and}
  \bibinfo{author}{\bibfnamefont{Y.}~\bibnamefont{Fukai}},
  \bibinfo{journal}{Phys. Rev. B} \textbf{\bibinfo{volume}{22}},
  \bibinfo{pages}{670} (\bibinfo{year}{1980}).

\bibitem[{\citenamefont{Schober and Stoneham}(1988)}]{sc88}
\bibinfo{author}{\bibfnamefont{H.~R.} \bibnamefont{Schober}} \bibnamefont{and}
  \bibinfo{author}{\bibfnamefont{A.~M.} \bibnamefont{Stoneham}},
  \bibinfo{journal}{Phys. Rev. Lett.} \textbf{\bibinfo{volume}{60}},
  \bibinfo{pages}{2307} (\bibinfo{year}{1988}).

\bibitem[{\citenamefont{Gillan}(1988)}]{gi88}
\bibinfo{author}{\bibfnamefont{M.~J.} \bibnamefont{Gillan}},
  \bibinfo{journal}{Phil. Mag. A} \textbf{\bibinfo{volume}{58}},
  \bibinfo{pages}{257} (\bibinfo{year}{1988}).

\bibitem[{\citenamefont{Noya et~al.}(1997)\citenamefont{Noya, Herrero, and
  Ram\'{\i}rez}}]{no97b}
\bibinfo{author}{\bibfnamefont{J.~C.} \bibnamefont{Noya}},
  \bibinfo{author}{\bibfnamefont{C.~P.} \bibnamefont{Herrero}},
  \bibnamefont{and}
  \bibinfo{author}{\bibfnamefont{R.}~\bibnamefont{Ram\'{\i}rez}},
  \bibinfo{journal}{Phys. Rev. Lett.} \textbf{\bibinfo{volume}{79}},
  \bibinfo{pages}{111} (\bibinfo{year}{1997}).

\bibitem[{\citenamefont{Herrero}(1997)}]{he97}
\bibinfo{author}{\bibfnamefont{C.~P.} \bibnamefont{Herrero}},
  \bibinfo{journal}{Phys. Rev. B} \textbf{\bibinfo{volume}{55}},
  \bibinfo{pages}{9235} (\bibinfo{year}{1997}).

\end{thebibliography}
\end{document}